\UseRawInputEncoding
\documentclass[aps,prx,twocolumn,amsmath,amssymb,superscriptaddress,floatfix]{revtex4-2}

\usepackage[english]{babel}
\makeatletter
\@ifundefined{l@en}{%
  \expandafter\let\csname l@en\endcsname\l@english
  \expandafter\let\csname captionsen\endcsname\captionsenglish
  \expandafter\let\csname dateen\endcsname\dateenglish
  \expandafter\let\csname extrasen\endcsname\extrasenglish
  \expandafter\let\csname noextrasen\endcsname\noextrasenglish
}{}
\makeatother
\usepackage[hidelinks]{hyperref}
\usepackage{graphicx}
\usepackage{multirow}
\usepackage{amsfonts}
\usepackage{amsthm}
\usepackage{mathrsfs}
\usepackage{xcolor}
\usepackage{booktabs}
\usepackage{ragged2e}
\usepackage{import}
\usepackage{pifont}
\usepackage[percent]{overpic}
\usepackage[normalem]{ulem}
\usepackage{tabularx}
\usepackage{siunitx}


\newcommand{\rlab}[1]{\makebox[1.2em][l]{#1}}
\newcommand{\brow}[4]{\textcolor{black}{#1} & \textcolor{black}{#2} & \textcolor{black}{#3} & \textcolor{black}{#4}}

\newcolumntype{Y}{>{\centering\arraybackslash}X}

\definecolor{forestgreen}{rgb}{0.13, 0.55, 0.13}

\newcommand{\new}[1]{\textcolor{black}{#1}}

\begin{document}

\title{Photonic Quantum-Accelerated Machine Learning}

\author{Markus Rambach}
\email{m.rambach@uq.edu.au}
\affiliation{Quantum Technology Laboratory, School of Mathematics and Physics, The University of Queensland, Brisbane, QLD 4072, Australia}

\author{Abhishek Roy}
\affiliation{Quantum Technology Laboratory, School of Mathematics and Physics, The University of Queensland, Brisbane, QLD 4072, Australia}
\affiliation{Department of Physics and Astronomy, Macquarie University, Sydney, NSW 2113, Australia}

\author{Alexei Gilchrist}
\affiliation{Department of Physics and Astronomy, Macquarie University, Sydney, NSW 2113, Australia}

\author{Akitada Sakurai}
\affiliation{Okinawa Institute of Science and Technology Graduate University, Onna-son, Okinawa 904-0495, Japan}

\author{William J. Munro}
\affiliation{Okinawa Institute of Science and Technology Graduate University, Onna-son, Okinawa 904-0495, Japan}

\author{Kae Nemoto}
\affiliation{Okinawa Institute of Science and Technology Graduate University, Onna-son, Okinawa 904-0495, Japan}

\author{Andrew G. White}
\affiliation{Quantum Technology Laboratory, School of Mathematics and Physics, The University of Queensland, Brisbane, QLD 4072, Australia}

\begin{abstract}
Machine learning is widely applied in modern society, but has yet to capitalise on the unique benefits offered by quantum resources.
Boson sampling---a quantum-interference based sampling protocol---is a resource that is classically hard to simulate and can be implemented on current quantum hardware.
Here, we present a quantum accelerator for classical machine learning, using boson sampling to provide a high-dimensional quantum fingerprint for reservoir computing.
We show robust performance improvements under various conditions: imperfect photon sources down to complete distinguishability; scenarios with severe class imbalances, classifying both handwritten digits and biomedical images; and sparse data, maintaining model accuracy with twenty times less training data.
\new{
Crucially, we demonstrate the acceleration of our scheme on a photonic quantum processing unit, providing experimental validation that boson-sampling-enhanced learning with Fock states delivers real performance gains on actual quantum hardware.
}
\end{abstract}

\maketitle

\section{Introduction}

Machine learning (ML) and quantum physics are two of the most transformative fields in the modern era of science and technology. ML has made significant advances in the last decade, especially in the last few years, and has firmly entered in the mainstream of every-day life. Quantum is not quite at the same point yet, but it is becoming more prominent to the public and increasingly integrated into emerging technologies, where it promises to advance our capabilities and open new pathways beyond classical limits. This leads to an obvious question: how can these two fields help each other? ML is already used widely to optimise quantum algorithms~\cite{nautrup_optimizing_2019,khairy_learning_2020,cheng_quantum_2024} and create or enhance quantum experiments~\cite{Krenn2016,melnikov_active_2018,strikis_learning-based_2021}, however, it has not yet tapped into the potential of real-world improvements that quantum offers in return. Here, we tackle this issue and investigate how quantum resources can augment or accelerate learning in verifiable and relevant ways, laying the groundwork for new quantum-classical hybrid computational models.

Recently, landmark quantum experiments in several physical platforms~\cite{Arute2019,zhong_quantum_2020,wu_strong_2021,madsen_quantum_2022,liu_robust_2025,abanin_observation_2025} have established computational tasks for which quantum systems outperform their best-known classical simulators. Unfortunately, these demonstrations---often relying on random circuit sampling or related approaches---rarely provide useful outputs outside of showing general advantage and in benchmarking contexts. In other words, while quantum processors can produce distributions inaccessible to classical algorithms within a reasonable time, those distributions are not typically aligned with any practical application.

The best-studied example is \emph{boson sampling}, a task in which $N$ indistinguishable single photons are injected into an $M$-mode interferometer and the output distribution is sampled~\cite{Aaronson2011,2013_broome_794,2013_crespi_545,2013_spring_798,2013_tillmann_540}. 
\new{This task is considered classically intractable: the output occupies a Hilbert space of size $\binom{M+N-1}{N}$, and each exact outcome probability is given by the permanent of a submatrix of the interferometer unitary, for which the best known classical algorithms scale as $\mathcal{O}(N \cdot 2^N)$~\cite{Clifford2018}.
So far this hardness has not been harnessed to advance other domains
directly. 
Computational hardness alone, however, does not imply superior ML feature extraction, but rather makes simulation of this type of resource expensive for small circuits and infeasible for large ones.}

In classical ML, reservoir computing~\cite{2009_lukosevicius_127,yan_emerging_2024}
is a framework where a fixed non-linear dynamical system (the reservoir) projects inputs into a high-dimensional space.
Then, only a simple readout layer is trained to extract useful features.
A crucial aspect of this framework is that the reservoir is static and not trained. Quantum reservoir computing~\cite{martinez-pena_dynamical_2021, bravo_quantum_2022, pfeffer_hybrid_2022, suzuki_natural_2022, martinez-pena_information_2023}---and its variants, quantum extreme learning
machine~\cite{suprano_experimental_2024, de_lorenzis_harnessing_2025}, and quantum extreme reservoir computing~\cite{sakurai_quantum_2022, hayashi_impact_2023}---extend reservoir computing to quantum systems, exploiting their naturally rich, high-dimensional and non-linear dynamics.
\new{Beyond photonics, the paradigm has been demonstrated on hardware at scale, most notably on a neutral-atom analog processor of up to 108 qubits applied to classification and time-series tasks~\cite{kornjaca2024}.}
Boson sampling can act as an intrinsically complex reservoir, creating a high-dimensional \textit{fingerprint} of the data. This can then be used as additional inputs by a classifier, e.g., for categorising images, 
a crucial task in autonomous driving and medical diagnostics.

\new{
Boson sampling has recently been connected to machine learning along several distinct lines, and it is worth separating them. 
On hardware, Gong et al.~\cite{gong_enhanced_2025} used a 8176-mode Gaussian boson sampler to enhance a perceptron on MNIST and Fashion-MNIST. 
There, the sampler runs once and each image enters only through classical mode selection, so the circuit is never reconfigured by the data. 
Cimini et al.~\cite{cimini_large-scale_2025} encoded each sample physically into a ${>}400$-mode frequency-multiplexed Gaussian boson sampler and showed that access to photon-number correlations, rather than mean fields alone, can be worth more than twenty percentage points.
On ten-class MNIST their reservoir reached ${\approx}60\%$ accuracy, below a single-layer linear classifier. 
In parallel, the role of quantum correlations has been probed directly: Di Bartolo et al.~\cite{DiBartolo2026} compared distinguishable and indistinguishable two-photon inputs in a reconfigurable integrated circuit with feedback and found a clear advantage for indistinguishable states on time-series forecasting, while Joly et al.~\cite{Joly2025}, working with photon pairs in a multimode fibre, found coincidence-based features superior to intensity but could not resolve a distinguishability effect at two photons. 
Numerical studies have mapped the state hierarchy in more detail~\cite{nerenberg_photon_2025} and extended boson sampling to quantum kernels and shallow networks~\cite{sharifian_hybrid_2025}.
}

A recent proposal, Quantum Optical Reservoir Computing (QORC)~\cite{sakurai_quantum_2025}, introduces a quantum reservoir to accelerate the performance of a linear classifier, achieving higher accuracies while retaining the same training constraint---only the single readout layer is trained.
This can be efficiently implemented on a quantum processing unit (QPU) consisting of an integrated photonic chip, Fock-state inputs, and single-photon bucket detectors without the need for number-resolving capabilities.
In practice, we see quantum acceleration through lower requirements on the size of the training dataset and the number of training epochs in all investigated scenarios.
QORC also outperforms small shallow neural networks in terms of accuracy, while incorporating the reservoir into such network architectures leads to further performance gain.

\begin{figure}[!t]
\centering
\includegraphics[width=.8\columnwidth]{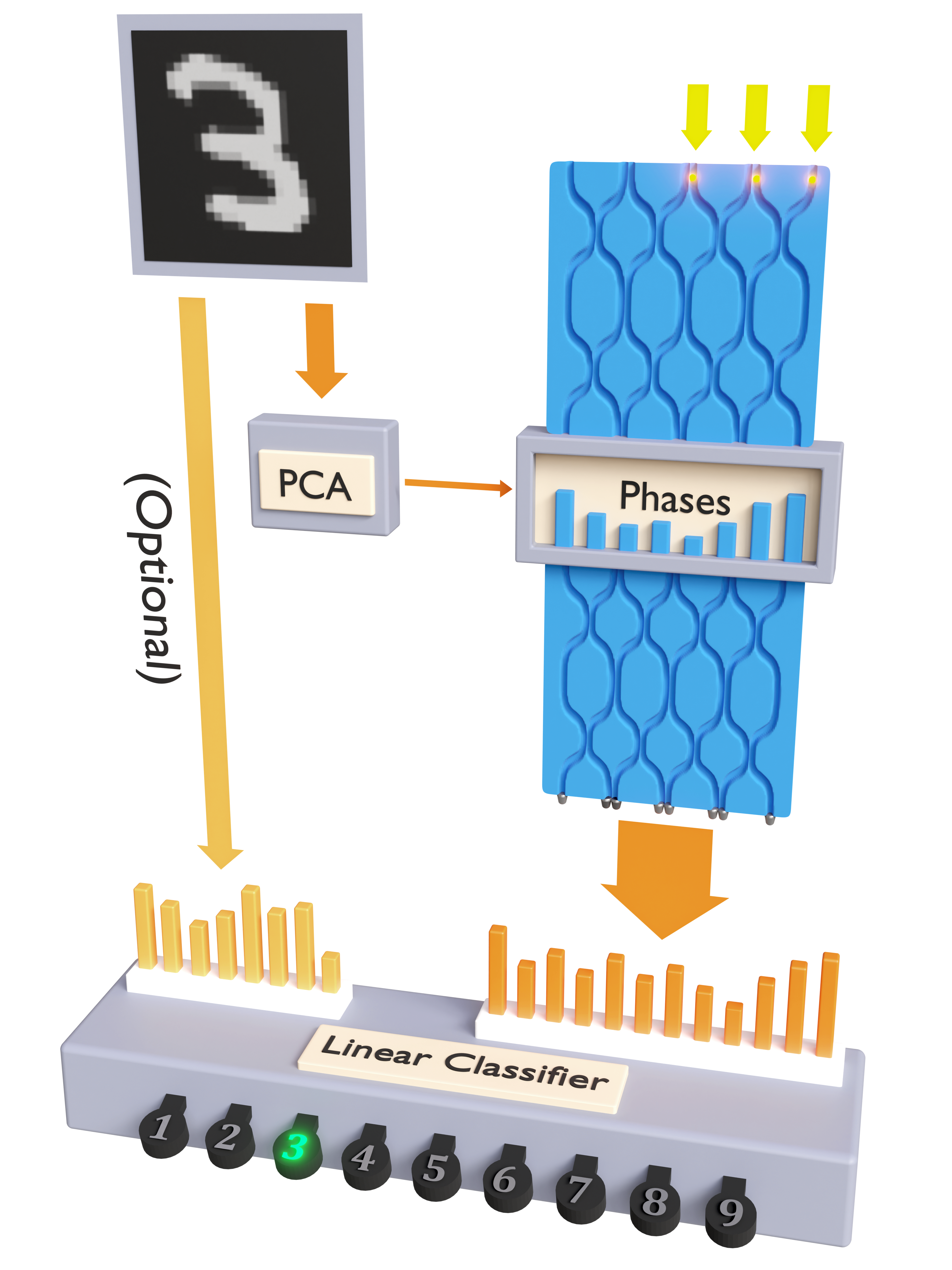}
\caption{\label{fig:scheme}
\justifying
Quantum Optical Reservoir Computing scheme.
An input image (top left) from a set of potential datasets is initially processed:
1) principal component analysis (PCA, right arrow), and 2) optional linearisation from 2D to 1D (left arrow).
The first $M$ components---matching the number of utilised modes---are forwarded to the boson sampling setup (thin arrow).
The setup consists of photons entering a pre-circuit (top right), then a layer where the PCA phases are encoded (centre), and finally the reservoir (bottom right).
The processed data is the input to the linear classifier (bottom), which finally finds the predicted classification label. \vspace{0mm}}
\end{figure}

Fig.~\ref{fig:scheme} illustrates the workflow of QORC.
We process data---in our case images---by first applying dimensionality reduction techniques to extract key features and map them onto phases in the quantum circuit. 
\new{In this paper we use principal component analysis, PCA, which is fitted on the training set only and then applied to the test set.
However, any reduction technique is suitable, e.g., wavelet compression would be another candidate.}
These phases can be imprinted on multi-photon quantum states---photon number $N$---via a set of phase shifters between two random interferometer circuits---\textit{pre-circuit} and \textit{reservoir}---of a photonic integrated chip.
The two random interferometers can differ, although for simplicity we have chosen them to be the same in all our experiments.

We measure the multi-mode output---mode number $M$---through single photon detection without number-resolving capabilities, to obtain a high-dimensional distribution of multi-photon coincidences. 
\new{Bucket detection restricts the accessible outcomes to one photon per mode, so the feature dimension is $D=\binom{M}{N}$ rather than the full Fock-space dimension $\binom{M+N-1}{N}$, setting the number of additional inputs quoted throughout.}
\new{The results are standardised and, when required, combined with the original classical features, then normalised by simply dividing by 255 for our 8-bit images (the same normalisation used in the non-accelerated classification), before being passed to the linear classifier.}
Unlike typical quantum machine learning models requiring optimisation or back propagation, QORC relies on random, fixed interferometers and uses the quantum evolution to non-linearly expand the feature space, enabling improved performance on non-linear classification tasks.

\new{
Our work is positioned differently from prior implementations of quantum reservoir computing. 
We keep the discrete-variable, Fock-state setting and encode each image as physical phases inside the interferometer, so that the quantum device is reconfigured per sample. 
Furthermore, rather than only competing on peak accuracy, we ask what a boson-sampling reservoir achieves under the conditions that decide whether such a scheme is deployable: imperfect sources, severe class imbalance, and scarce labelled data, with claims anchored to measurements on a photonic QPU.
}

\begin{figure}[t]
\centering
\begin{overpic}[width=\columnwidth]{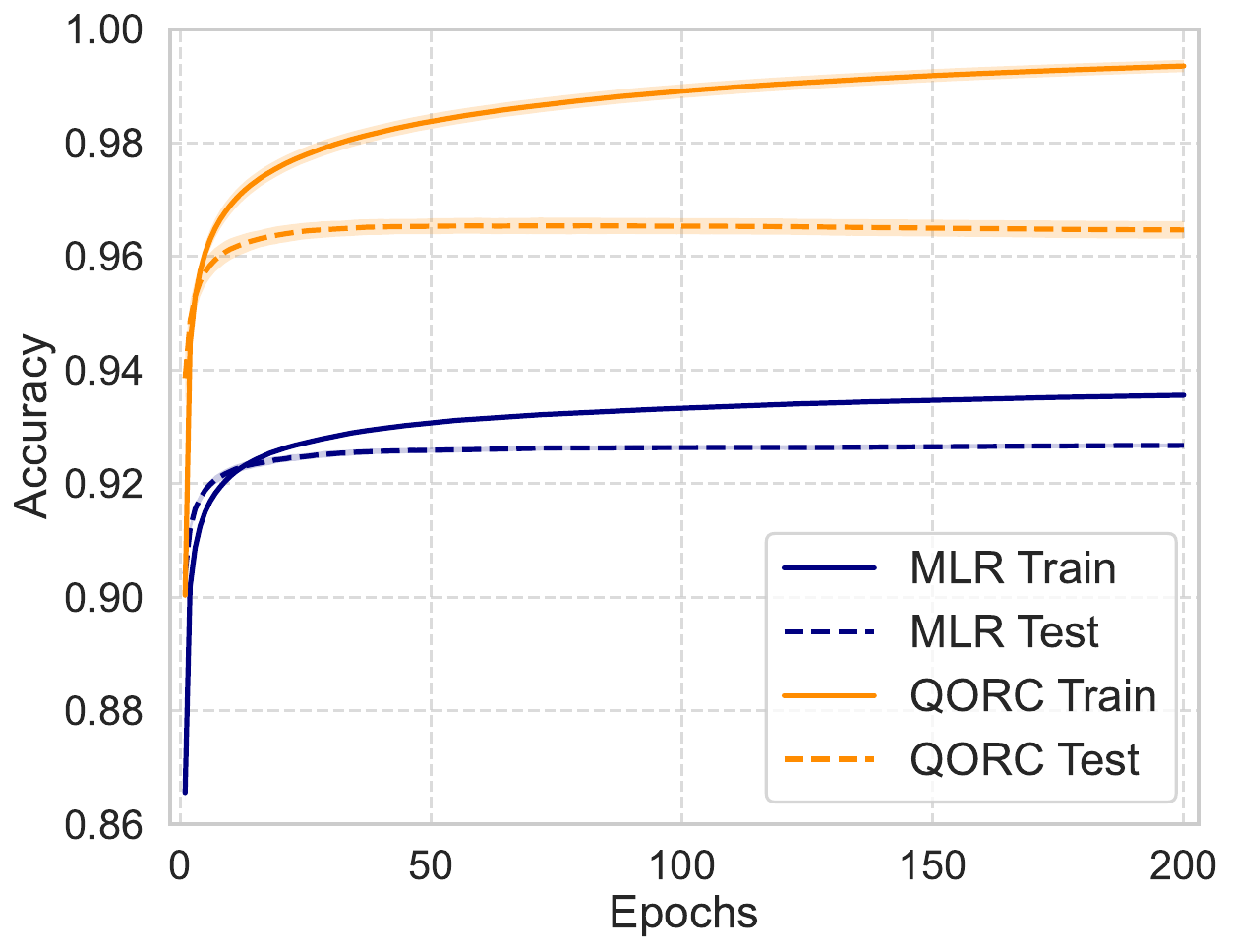}
    \put(2,80){\large\textbf{(a)}}
\end{overpic}
\hfill
\vspace{2mm}
\begin{overpic}[width=\columnwidth]{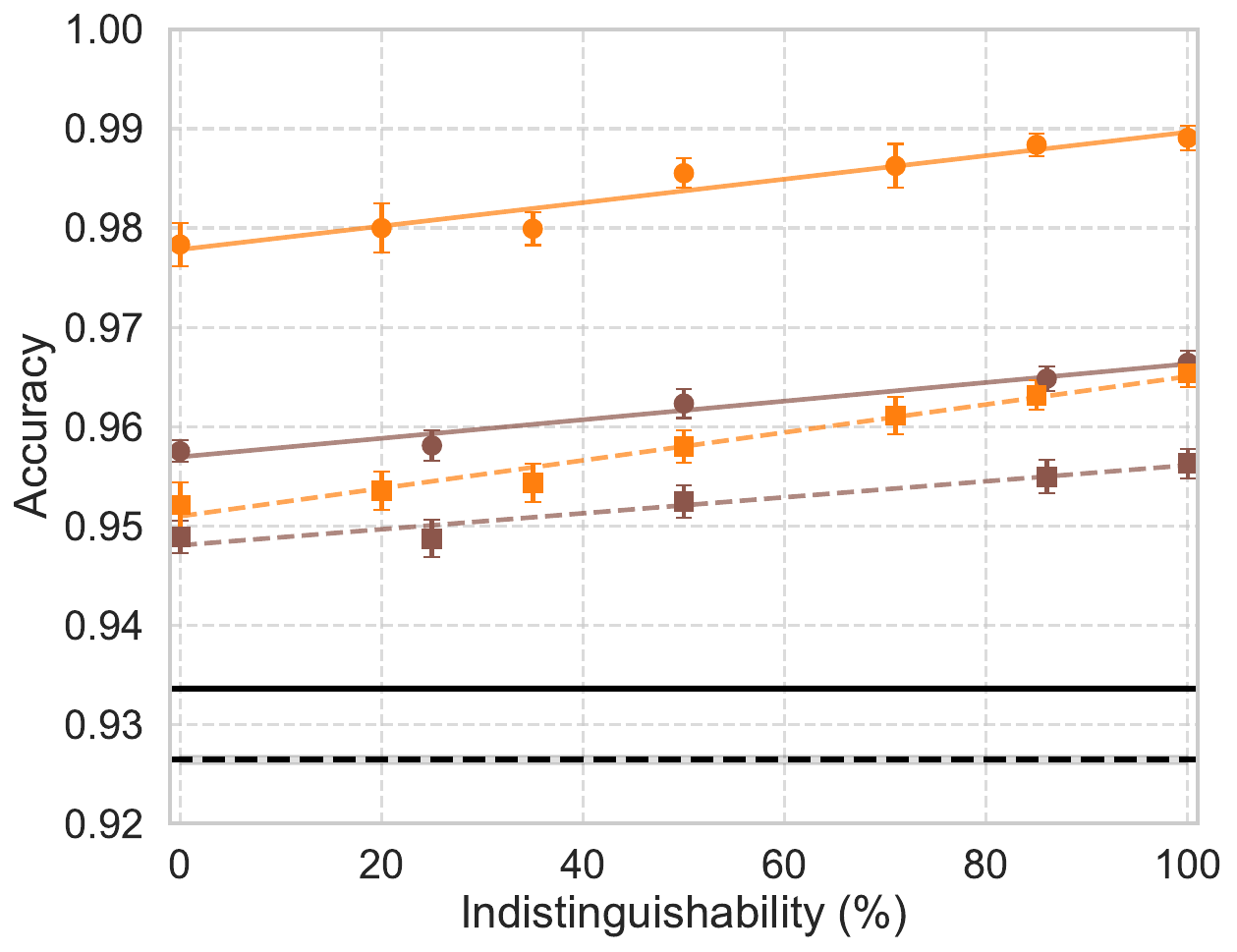}
    \put(2,80){\large\textbf{(b)}}
\end{overpic}
\caption{\justifying
    MNIST classification accuracies.
    QORC in orange ($N{=}3, M{=}20$) and brown ($N{=}3, M{=}12$), MLR in blue/black.
    Training (circles, solid lines) on 60,000 images, testing (squares, dashed lines) on 10,000 images.
    For each image, the reservoir is 30,000 samples.
    \new{Shaded region/error bars are the standard deviations of 100 classification runs: 10 different unitaries with 10 different train/test splittings.}
    (a) Accuracy as a function of training epochs.
    QORC outperforms the best value of MLR from the very first epoch.
    \new{Best test performance on both models is above 50 epochs.
    Shaded region is very small and only visible for QORC Test.}
    (b) Accuracy as a function of single photon indistinguishability.
    Coloured fitted lines to guide the eye only.
    All accuracies for all indistinguishabilities of either reservoir size here surpass the maximal achievable training accuracy of MLR.}
\label{fig:MNIST_combined}
\end{figure}

\section{Results}

To evaluate the computational potential of QORC, we examine its ability to accelerate image classification tasks across a range of configurations. 
We numerically investigate a range of configurations, one to five photons, $N{=}$1-5, into $M{=}12/20/24$ optical modes; experimentally, we use $N{=}3$ and $M{=}12$.
This is informed by various factors: 
1)~the available QPUs on the Quandela cloud architecture, which offer either 12 (Ascella) or 24 (Belenos) modes; 
2)~reasonable runtimes for the necessary number of samples on the QPU and hence a combination of the photon number and internal losses in the cloud system; 
3)~already available numerical simulations from our prior work~\cite{sakurai_quantum_2025}; 
4)~reasonable runtimes on the local numerical simulations; and 
5)~fulfilling the boson sampling condition of ideally keeping $M{>}N^2$.
All simulations assume noiseless operation, i.e., they are performed with ideal sources---perfect multi-photon suppression $g^{(2)} {=} 0$, and ideal quantum indistinguishability $\mathcal{I}{=}1$---and lossless QPUs, unless stated otherwise. 
\new{Quoted uncertainties are standard deviations over 100 classification runs, comprising 10 random reservoir unitaries $\times$ 10 train/test
splittings. 
What dominates the variance is dependent on the dataset: for almost all investigated cases, it is the choice of unitary rather than by the data splitting, with an approximate 7:1 ratio.
The exception is DermaMNIST, which is smaller and heavily imbalanced. 
There, the splitting dominates, as training instabilities outweigh the contribution from the reservoir.}

The model is built in Keras, a high-level neural network interface based on TensorFlow for computation.
It is constructed as a linear sequence of an input layer of variable size---dependent on whether a reservoir is used and its size---and an output layer for the possible outcomes/classes.
The model is then optimised using the AdaGrad algorithm with a learning rate $\eta{=}0.05$, a batch size of 128, and 100 epochs, unless stated otherwise.

\begin{table}[t]
\caption{\label{tab:MNIST}
\justifying
MNIST linear classification (MLR) results for different input data: i) original input data; ii) our scheme (QORC); iii) using the reservoir only; iv) adding the PCA to the original data; \new{v) Random Fourier Features (RFF); and vi) an Extreme Learning Machine (ELM).}
Training on 60,000 images for 100 epochs, testing on 10,000 images.
\new{Errors in brackets are the standard deviation of 100 classification runs: 10 different unitaries with 10 different train/test splittings.}
\emph{top}: 3 photons into 20 modes (1,140 additional inputs).
\emph{bottom}: 5 photons into 24 modes (42,504 additional inputs).
$\Delta$MLR is the difference from the input of the original data only.
}
\begin{tabular*}{\columnwidth}{l@{\extracolsep{\fill}}ccc}
\toprule
& \multicolumn{3}{c}{Accuracy (\%)}  \\
\cmidrule(lr){2-4}
Input Data & {Train} & {Test} & {$\Delta$MLR} \\
\midrule
$N=3, M=20$ & & & \\
\cmidrule(r){1-1}
\rlab{i)} Original data     & 93.33 (1) & 92.64 (4) & {\text{--}}  \\
\rlab{ii)} QORC             & 98.9 (1)  & 96.5 (1)  & 3.9 \\
\rlab{iii)} Reservoir only  & 97.7 (2)  & 95.6 (2)  & 3.0 \\
\rlab{iv)} Original + PCA   & 99.32 (2) & 92.61 (4) & 0.0 \\
\brow{\rlab{v)} RFF}{99.3 (1)}{96.7 (1)}{4.1} \\
\brow{\rlab{vi)} ELM}{95.7 (3)}{93.8 (3)}{1.2} \\
\midrule
$N=5, M=24$ & & & \\
\cmidrule(r){1-1}
\rlab{i)} Original data     & 93.33 (1) & 92.64 (4) & {\text{--}}  \\
\rlab{ii)} QORC             & 100 (0)   & 97.5 (1)  & 4.9 \\
\rlab{iii)} Reservoir only  & 100 (0)   & 97.5 (1)  & 4.9  \\
\rlab{iv)} Original + PCA   & 99.32 (2) & 92.61 (4) & 0.0 \\
\brow{\rlab{v)} RFF}{99.7 (1)}{97.5 (1)}{4.9} \\
\brow{\rlab{vi)} ELM}{96.8 (4)}{95.2 (4)}{2.6} \\
\bottomrule
\end{tabular*}
\end{table}

\subsection{Performance with balanced datasets}

We investigate two inherently balanced datasets: in the main paper the MNIST dataset~\cite{deng_mnist_2012}, 70,000 handwritten digits as 28-by-28 pixel greyscale images; and in the Supplementary Material Sec.~\ref{sec:digits}~\cite{Supp},
the smaller DIGITS data set~\cite{e_alpaydin_pen-based_1996}, 1797 images of 8-by-8 handwritten digits.
\new{Unless stated otherwise, we compare our scheme with multinomial logistic regression (MLR)---a single dense layer with softmax activation over the ten output classes, trained with categorical cross-entropy---using the same data preprocessing pipeline.
We adopt this baseline because the computational cost of both the MLR and our scheme scales linearly with dataset size.
The reservoir features for QORC are moreover measured, not computed: a single optical pass, whereas classical reservoirs with the same linear cost scaling must evaluate all $D = \binom{M}{N}$ values explicitly.}

Fig.~\ref{fig:MNIST_combined}(a)
shows the performance of QORC on classifying the MNIST dataset~\cite{deng_mnist_2012} as a function of the number of training epochs for a simulated quantum reservoir with three photons, $N{=}3$, into twenty modes, $M{=}20$.
It highlights the performance improvements of the QORC (orange) over the non-accelerated classification (blue).
In both cases, the training accuracy (solid lines) is above the test accuracy (dashed lines) as expected.
\new{Since test accuracies plateau after around 50 epochs, we report results for 100 epochs from here on to ensure we don't miss late improvements.}
We can also see that our scheme outperforms the best value of MLR from the very first epoch, highlighting the quantum acceleration in the number of required training runs.
See Supplemental Material Sec.~\ref{sec:confusion}~\cite{Supp} for more details on how we derive the accuracies.

We furthermore compare the performance of the classifier under different inputs: i) multinomial logistic regression (MLR) on the original data; ii) original data plus reservoir (QORC); iii) reservoir only~\cite{sakurai_simple_2025}; iv) original data plus PCA; \new{v) Random Fourier Features (RFF); and vi) an Extreme Learning Machine (ELM).}
Table~\ref{tab:MNIST} shows the results: for ii) we see an increase in test accuracy of 3.9\% if we add the reservoir to the original data for a reservoir of $N{=}3, M{=}20$, which is 1,140 additional inputs.
Using the reservoir only, iii), the advantage shrinks to a still significant 3.0\%, and finally, if we add the PCA to the original input, iv), the advantage disappears as expected, as the linear PCA simply gets incorporated in the MLR network.
Similarly, if we move to $N{=}5, M{=}24$---an additional 42,504 inputs to the network---the advantage jumps to 4.9\%.
This is independent of whether the original MNIST data is added or not, as the reservoir now incorporates all relevant features for learning.
Again, adding the PCA yields no advantage.
Further analysis of the effect of the number of photons on the model accuracy for a fixed number of modes $M{=}20$ and the reproducibility of results for different unitaries and input states can be found in the Supplemental Material Sections~\ref{sec:reproducibility} and \ref{sec:M20}~\cite{Supp}.

\new{To benchmark QORC against classical methods that also rely on fixed, random nonlinear feature maps, we additionally compare to v) RFF~\cite{Rahimi2007} and vi) an ELM~\cite{Huang2006}. 
RFF approximates a kernel expansion by projecting the input through random Gaussian weights followed by a sinusoidal nonlinearity. 
ELM randomly initialises and fixes the weights of a single hidden layer and trains only the output layer, directly paralleling QORC's fixed reservoir with a trainable readout. 
As shown in Table~\ref{tab:MNIST}, RFF reaches a similar improvement to QORC, while ELM yields a more modest gain, clearly below our scheme. 
Implementation details and additional comparisons of our scheme to RFF and ELM are given in Supplemental Material Section~\ref{sec:RFF_ELM}~\cite{Supp}.}

\subsection{First-order vs second-order quantum coherence}

The performance of the interferometric circuits---\textit{pre-circuit} and the \textit{reservoir}~\cite{sakurai_quantum_2025}---for the boson sampling strongly depends on the quality of the single photon Fock states.
We quantify the quality through multi-photon suppression and indistinguishability~\cite{somaschi_near-optimal_2016}.
The latter plays an especially crucial role, as it effectively tunes the system between regimes dominated by first- and second-order quantum coherence.
Completely distinguishable photons exhibit no intensity correlations with one another, yet retain coherence in their field amplitude and can therefore interfere with themselves.
In Fig.~\ref{fig:MNIST_combined}(b)
we fix the multi-photon suppression at the experimentally achieved value, $g^{(2)}$=1.95\%, and vary the HOM indistinguishability $\mathcal{I}$,
for the quantum reservoirs $N{=}3~\&~M{=}12$ (brown), $N{=}3~\&~M{=}20$ (orange), and the original data (black).
Again, the training accuracies are represented by circles and solid lines while the test accuracies are squares and dashed lines.

We observe that higher indistinguishability is correlated with higher training accuracy: the stronger the multi-photon interference, the more expressive the feature map and the finer the quantum fingerprint.
Even for fully distinguishable photons ($\mathcal{I}{=}0$), QORC remains advantageous due to first-order quantum coherence.
While the Hilbert-space dimensionality is reduced compared to the entangled regime ($\mathcal{I}{>}0$), QORC nonetheless provides additional informational capacity that improves classification.
Naturally, the larger quantum reservoir performs better, however, a large improvement over 2\% between different reservoirs is only found in the training accuracies, not in testing where the improvement is below 1\%.

\begin{figure}[t!]
\centering
\includegraphics[width=\linewidth]{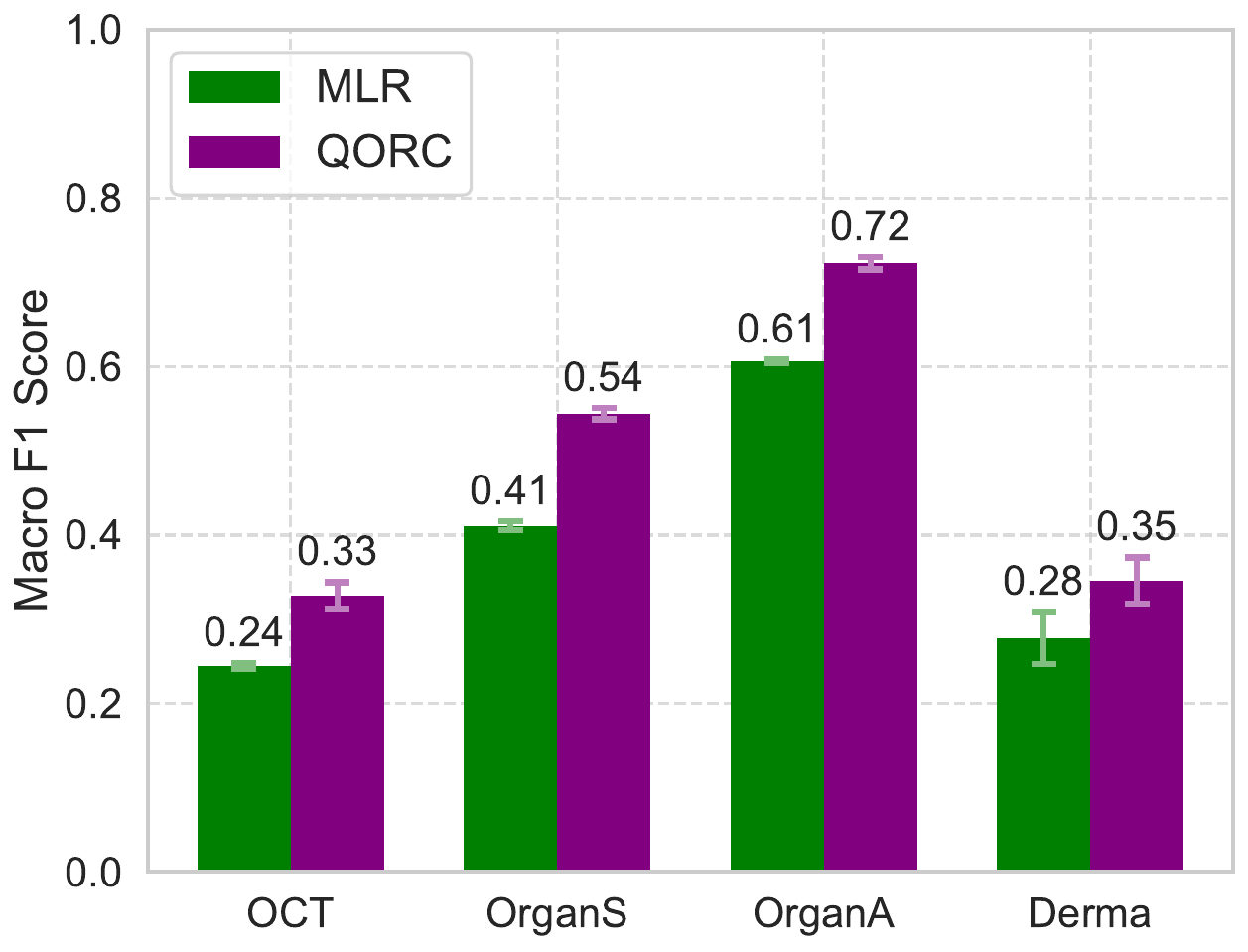}
\caption{\label{fig:MedMNISTF1}
\justifying
QORC classification F1 score improvements on different datasets in MedMNISTv2 for 200 training epochs.
Quantum reservoir: $N{=}3, M{=}20$. For each image, the reservoir is 30,000
samples.
OCT, OrganS, and OrganA are all greyscale images, Derma is in colour.
\new{Error bars are the standard deviations of 100 classification runs: 10 different unitaries with 10 different train/test splittings.}
See Supplemental Material Sec.~\ref{sec:imbalanced}~\cite{Supp}
for more details on the datasets.
}
\end{figure}

\begin{figure*}[!t]
\centering
\includegraphics[width=2\columnwidth]{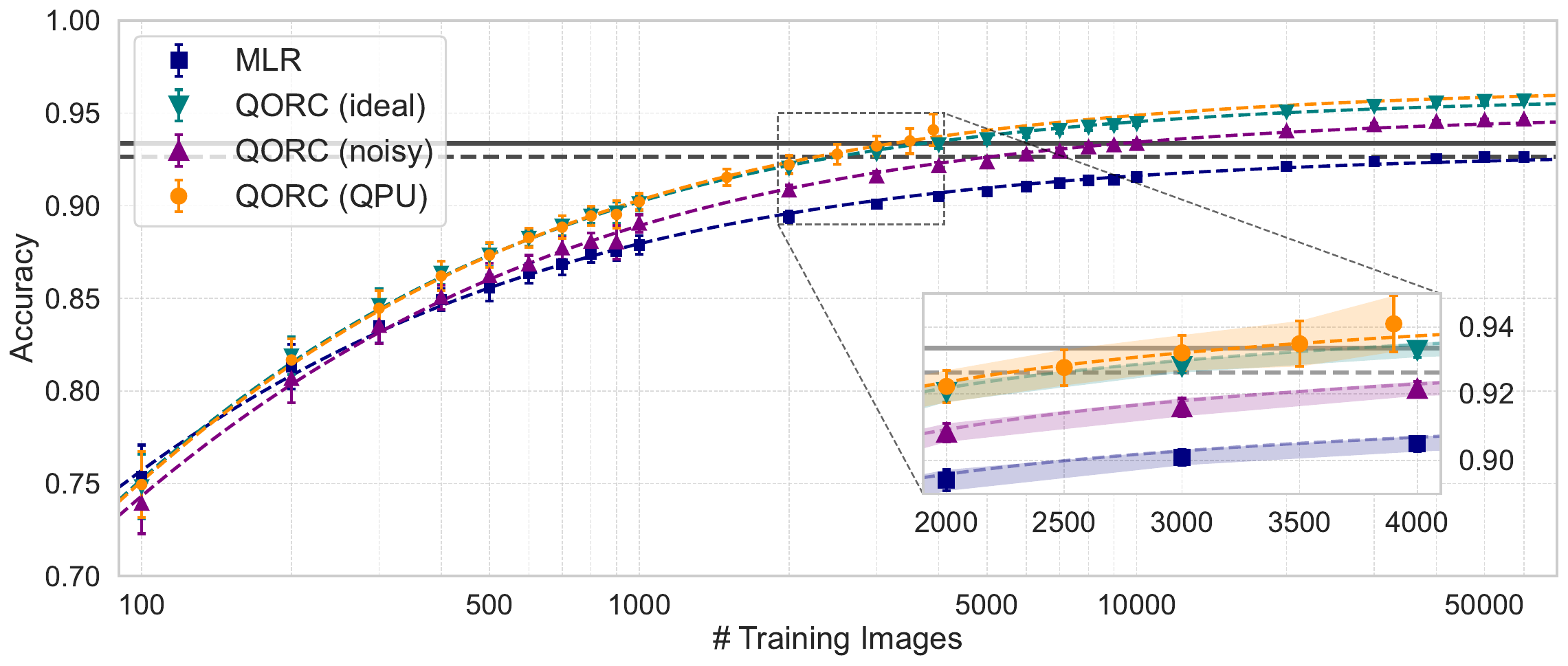}
\vspace{-3mm}
\caption{\label{fig:MNIST_ntrain}
\justifying
MNIST classification accuracy dependent on (balanced) training dataset size $n_{tr}$. 
MLR in blue squares, QORC with $N{=}3, M{=}12$ in teal triangles (ideal, no noise), purple inverted triangles (noisy, $g^{(2)}{=}1.95\%$, $\mathcal{I}{=}0$), and orange circles (QPU, $g^{(2)}{=}1.95\%$, $\mathcal{I}{=}86.36\%$).
Error bars are the standard deviation on 50 subsets of training data: too small to see for large numbers of training images, but significant for smaller $n_{tr}$ and the QPU data.
Grey lines show the best possible performances for train (solid) and test (dashed) accuracies with MLR.
Training for 100 epochs, testing (dashed lines) on $n_{te}{=}$10,000 images apart from QPU data, where $n_{te}=4669{-}n_{tr}$.
Main: Dashed coloured lines to guide the eye only, Hill function fit: $L / (1 + (x_{50} / x)^k)$ with $(L, k, x_{50}) = (0.929, 0.61,  8.6), (0.959, 0.64, 13.3), (0.950, 0.61, 13.2), (0.964, 0.62, 13.1)$ for MLR and QORC (ideal, noisy, QPU), respectively.
Inset: Shaded regions indicate $1\sigma$ band. 
No QPU data available beyond $n_{tr}=3900$.
}
\vspace{-5mm}
\end{figure*}

\subsection{Performance with artificially imbalanced datasets}

Real-world data is often imbalanced---e.g., in biomedical imaging or fraud detection---where many more samples are available in common classes---healthy people or non-fraudulent transactions---compared to rare but critical ones---diseases or security risks.
These two examples show that imbalance is not a mere data collection artefact but rather reflects structural characteristics.
Models often achieve good performance in majority classes, but struggle on minority classes, leading to poor classification of cases with larger significance~\cite{gao_comprehensive_2025}.

Here, accuracy as a metric for performance can be misleading, as a naive classifier that simply predicts the favoured majority class can achieve deceptively high accuracies.
Hence, we move to a different criterion---the macro F1 score---to assess the performance of our models.
The macro F1 score is a simple metric that computes the harmonic mean of precision and recall for each class and then averages across all classes, weighing all components the same and hence taking class imbalance into account.
A value close to \textit{one} indicates the model performs well across all classes, a low value close to \textit{zero} hints at poor model performance on one or multiple classes, even if the overall accuracy is still high.

We start by artificially creating two imbalanced training datasets of 10,000 images derived from MNIST and evaluate the performance of QORC ($N{=}3, M{=}20$).
The first one applies a Gaussian filter with $\mu{=}4.75, \sigma{=}2$ to the training set, i.e., the number \textit{five} will be the majority (20\%) and the number \textit{zero} the smallest class (1.2\%).
Similarly, the second more severely imbalanced dataset has a majority (minority) class of 64\% (0.6\%) which are then randomly distributed across the 10 categories for each new training trial.
For a fair comparison, we also reduced the size of the balanced MNIST training data to match.
Individual images for training are chosen randomly for each trial to determine the mean F1 and its standard deviation.
We can see that QORC still improves the F1 score for all datasets on both training and test images, demonstrating its significant value for real-world applications.
Please see Supplemental Material Sec.~\ref{sec:imbalanced}~\cite{Supp} for further details on accuracies, F1 scores, and imbalances for the artificial datasets.

\subsection{Performance with naturally imbalanced datasets}

Biomedical diagnostics represent an important challenge to ML due to their often largely class-imbalance datasets which occur frequently in real-world applications. Such datasets---with one dominant class and significantly fewer images of other diagnostically more relevant classes---stress test many models as they tend to be biased towards majority classes, and underperform on rare but clinically critical cases~\cite{johnson_survey_2019}.

Accordingly, we explore the performance on {MedMNISTv2}~\cite{yang_medmnist_2023}, a variety of biomedical image datasets in greyscale and colour, all standardised to 28-by-28 pixels as well.
This large-scale collection of lightweight biomedical image datasets is specifically designed for benchmarking machine learning algorithms across diverse medical imaging techniques and tasks.
The performance of any classifier on MedMNISTv2 is significantly worse compared to standard MNIST as the images are more complex and varied, the signal-to-noise is lower, and generally the classes are not visually obvious and labels might be imperfect.
Our results are compared in Fig.~\ref{fig:MedMNISTF1}.
Despite the macro F1 scores being significantly lower compared to the artificially imbalanced datasets, QORC still yields improvement in all investigated scenarios.

\new{The significant improvement on both imbalanced and more complex datasets is a very promising, albeit preliminary, sign of QORC’s capability to extend beyond simple benchmarks to challenging real-world tasks.}
Our results suggest that QORC can effectively adapt to diverse data distributions and maintain strong performance beyond controlled benchmark settings.

\subsection{Training advantage}

Seeing that QORC exceeds the performance of MLR for any input photon number---see Supplemental Material Sec.~\ref{sec:M20}~\cite{Supp} for details on different $N$ for $M{=}20$---and all investigated datasets, the crossover point of the performance in terms of training data requirements becomes of interest.
It is crucial for ML models to still perform well with limited training data, as data collection is often costly and impractical in real-world scenarios.
Additionally, demonstrated robustness can hint at strong generalisation capabilities, with the model capable of learning meaningful patterns rather than noise.

Fig.~\ref{fig:MNIST_ntrain} illustrates the performance differences on random, balanced subsets of training data on MNIST.
We can see that QORC surpasses the maximally possible testing accuracy of MLR (black dashed) at around 2,500 training images for a perfect noiseless single photon source (teal triangles), a significant improvement of $(23.5\substack{+0.5 \\ -3.5})$ times compared to the 60,000 images used to reach peak performance in the MLR.
In the case of completely distinguishable photons (purple inverted triangles), the advantage shrinks to $11.8\substack{+0.2 \\ -1.8}$ times, i.e., only 5,000--6,000 training images are required for similar accuracy.
Finally, we collected distribution samples for 4,669 MNIST images from a physical QPU (Quandela Ascella, $g^{(2)}{=}1.95\%$, $\mathcal{I}{=}86.36\%$) for QORC (orange circles).
We observe performance of $(26.5\substack{+3.5 \\ -6.0})$ times training size reduction, marginally above the simulations for a perfect photon source but overlapping within error bars.
We attribute the slight discrepancy to the sampling nature of our experiments and simulations.
This close to ideal performance of the physical QPU is also supported by Fig.~\ref{fig:MNIST_combined}(b),
where the accuracy of an ideal and real source (rightmost two brown squares) are close together, and far above the distinguishable photons (leftmost brown square).
This QPU-confirmed quantum acceleration has multiple \new{potential} implications for real-world applications: 
\new{compared to linear classifiers, we observe }
1) lower training costs, computationally and in terms of data labelling efforts; 
2) larger independence from data biases; and
3) applicability in many domains that naturally have sparse data.

\begin{figure*}[t]
\centering
\includegraphics[width=2\columnwidth]{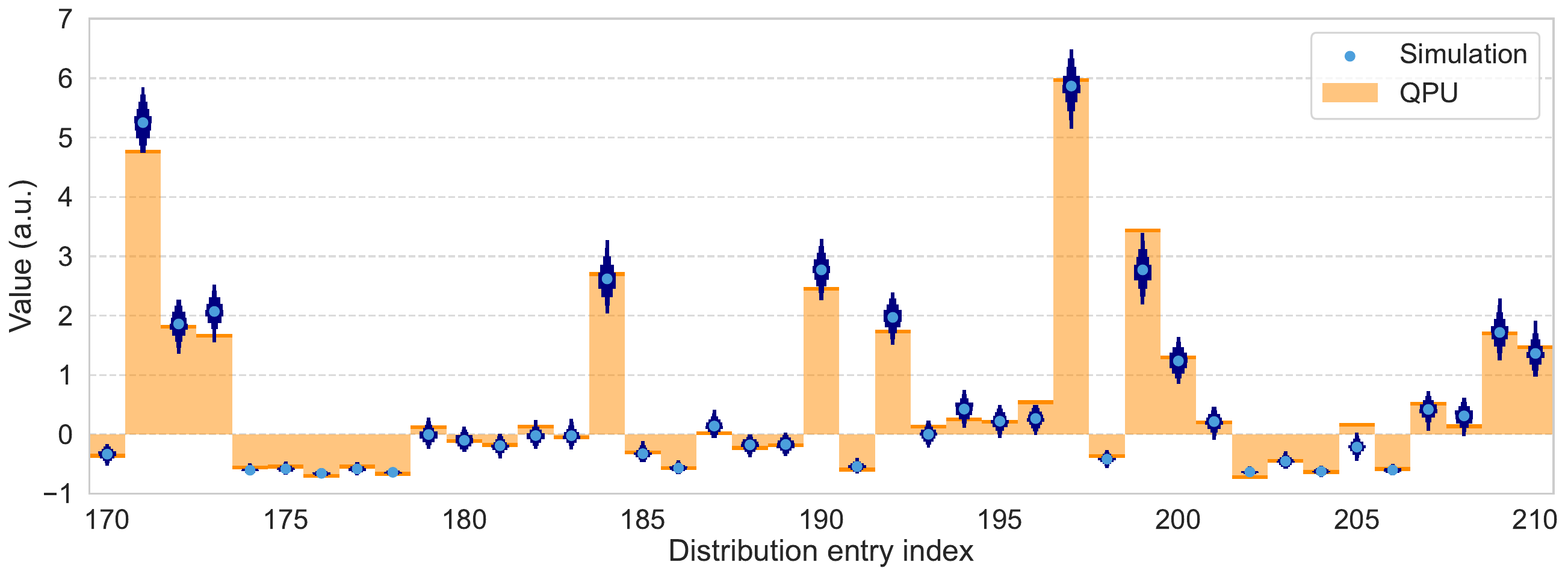}
\vspace{-3mm}
\caption{\label{fig:quandela}
\justifying
Comparison of experimental photonic (orange) reservoir distributions with 500 numerical simulations (blue) for the same noise for $N{=}3, M{=}12$ (total of 220 distribution entries).
Light blue dots are the means of the 500 numerical simulations, error bars (dark blue) have a thickness proportional to the occurrence of certain values, with top and bottom the minima and maxima.
We see a very high overlap between numerical simulations and boson sampling on a quantum cloud platform, qualitatively and quantitatively, even for the moderate 30,000 collected samples per image.
}
\vspace{-3mm}
\end{figure*}

\subsection{Quantum processor performance}

Finally, we investigate the performance on Quandela's Ascella QPU in more detail, i.e., how closely the distributions from numerical simulations overlap with the photon counting results on the Quandela platform.
For a straight comparison, we simulate the experimental values for $g^{(2)}$ and indistinguishability for the Ascella QPU using three photons into 12 modes.
We pick the subset of 4669 images that we processed on the QPU and compare the obtained distributions using different metrics.
Firstly, we renormalise the distribution so that the sum equals \textit{one} and all values are positive. 
In that way, we can use standard distribution comparison metrics, and find that the simulation of the experiment and the actual experiment are indeed very similar, with e.g. total variational distance (TVD) $=0.097{\pm}0.009$ or Jensen-Shannon (JS) distance $=0.091{\pm}0.007$, both ideally \textit{zero} for identical and \textit{one} for maximally different distributions.
These are encouraging results, but due to the initial re-normalisation, the compared probabilities are not exactly the same as the input into our classifier.
Secondly, we use the standardised input distribution for comparison. 
This distribution is not normalised or strictly positive, it stems from a standardisation to \textit{zero} mean and \textit{unit} variance.
Here, we chose the Kolmogorov-Smirnov (KS) statistic 
as a metric because TVD and JS distance are not applicable.
Our value KS $=0.064{\pm}0.023$ is close to \textit{zero}, which would be the ideal case for two perfectly overlapping distributions.
\new{
For comparison, two independent draws from the same simulated distribution give TVD${=}0.053(5)$, JS${=}0.054(4)$, and KS${=}0.044(9)$, the floors set by finite sampling alone.
The experiment exceeds this floor by factors of 1.8, 1.7, and 1.5, leaving a small but statistically significant residual, which we attribute to device imperfections not captured by our noise model.
}

Unfortunately, the above numbers alone cannot capture all the nuances of how two distributions differ.
For example, two distributions might have the same KS statistic but differ in shape, skew, or where the differences occur.
Plotting the distributions side by side, however, allows the human eye to see these subtleties.
Visualisation provides context that a single number might hide, making it easier to understand the practical significance of the differences.
Fig.~\ref{fig:quandela} shows such a plot of differences for a subset of distribution entries on one randomly chosen image from MNIST.
The photonic experiment on a QPU (orange) is compared to 500 numerical simulations for the same noise (light blue dot is its mean), with the thickness of the error bars corresponding to a histogram of value occurrences.
This showcases the high qualitative and quantitative overlap between numerical and photonic experiments, with the obtained \textit{fingerprint} of the image nearly identical within errors.

\subsection{QORC vs deep networks}

Following these very encouraging results on the achievable accuracies for different training datasets and significantly lower requirements on the number of training images, we explore the potential of QORC beyond linear classifiers on the MNIST dataset.
In order to do this in a fair way, we need to optimise the hyperparameters of the networks under consideration.
To find the best architectures for different network models, we employ KerasTuner~\cite{omalley2019kerastuner}, a scalable hyperparameter optimisation framework within Keras.
We compared four models: the MLR, a shallow network with a fixed or variable number of units, and a deep neural network.
We iterate the optimisation over 1,000 potential models trained for 30 epochs each and consider the reservoir with $N{=}3, M{=}20$.
Please see the Supplemental Material Sec.~\ref{sec:KerasTuner}~\cite{Supp} for further details on the procedure and the searched parameter space.

\begin{figure}[t]
\centering
\includegraphics[width=\columnwidth]{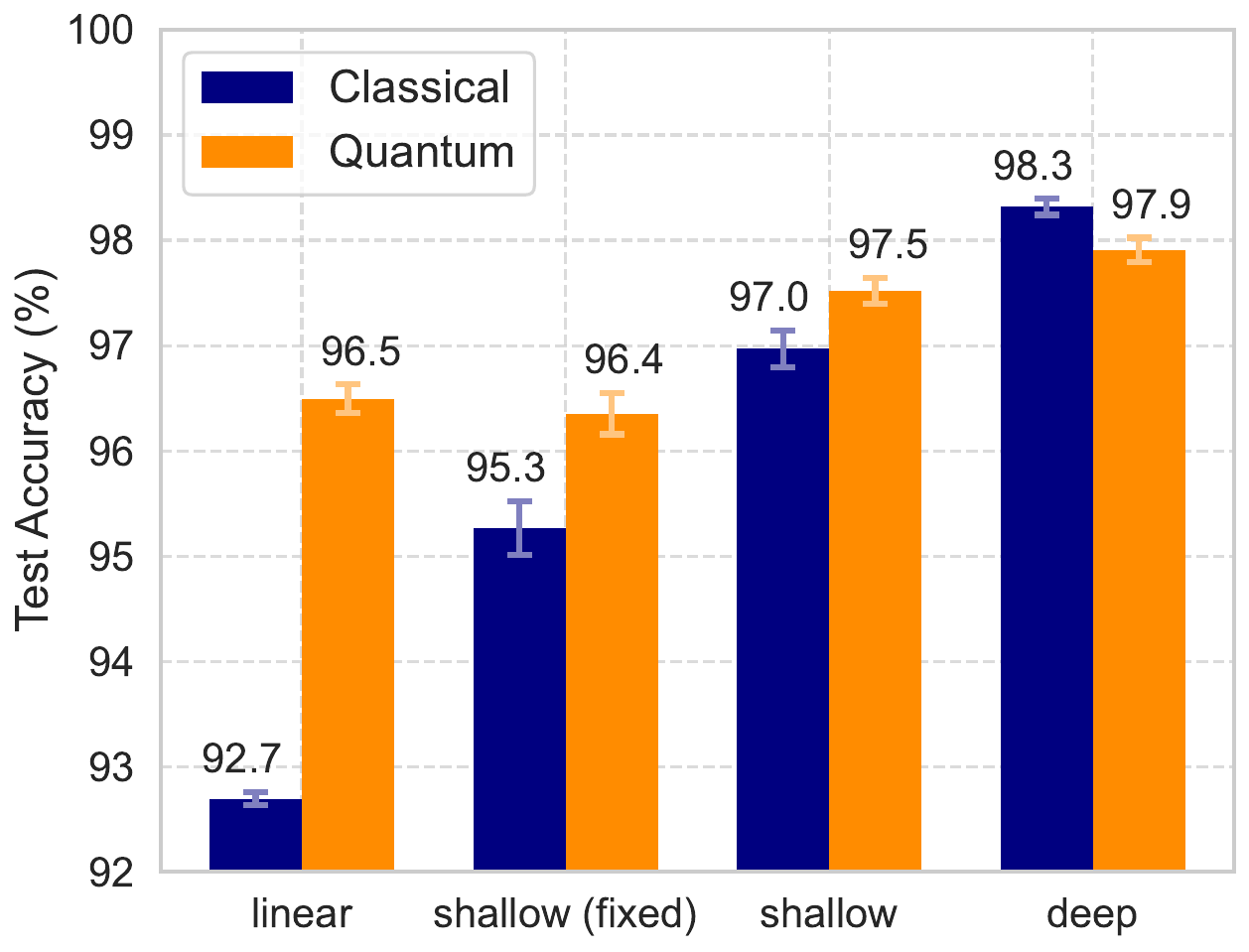}
\caption{\label{fig:MNIST_networks}
\justifying
MNIST classification accuracy for different network architectures.
QORC for $N{=}3, M{=}20$.
For each image, the reservoir is 30,000
samples.
Optimal network parameters found by KerasTuner, iterating 1,000 potential models with 30 epochs each.
Model search space parameters in Supplementary Material Sec.~\ref{sec:KerasTuner}~\cite{Supp}
MNIST data with: Blue, no reservoir; Orange, quantum reservoir.
\new{Error bars are the standard deviations of 100 classification runs: 10 different unitaries with 10 different train/test splittings for quantum, 100 different train/test splittings for classical.}}
\end{figure}

Fig.~\ref{fig:MNIST_networks} summarises the results and shows that the improvement of QORC over the non-accelerated neural networks reduces with network complexity.
Adding even a modest-sized reservoir seems to be disadvantageous compared to the original input data for the optimised deep neural network with three hidden layers.
We attribute this to an increase in inputs with very noisy or no actual information, i.e., many of the values in the boson sampling distribution are close to zero and prone to under-sampling when using a maximal sample size of 30,000.
This could be avoided by either using a larger number of samples or by identifying more important features in the distribution and feeding only those to the network.
Either way, we are not expecting QORC to surpass the performance of a deep neural network as this architecture already offers rich modelling capabilities of non-linear behaviour.
We also believe the overall behaviour of reduced improvements by QORC is due to the increased capabilities of the networks to go beyond linear classification.
Note that as the network complexity increases, QORC still achieves progressively higher accuracies.

\vspace{-3mm}
\section{Discussion}
\vspace{-1mm}

We are currently in a period of exponential improvement in quantum photonics, as the field leverages the capabilities of foundry fabrication~\cite{alexander_manufacturable_2025}. State of the art commercial systems have a single-channel transmittance of $\eta_{1} {\sim} 0.22$--$3.04\%$---which combines circuit throughput and detector efficiency, but only captures partial aspects of the source efficiency. With typical clock rates of 80~MHz, the five-photon rate~\footnote{$R_{tot} = \alpha * RR / N * \eta_{1}^{N} * FF * PF$, with $N$, the number of photons, $\alpha \approx 1/N$ and $FF * PF=0.56$, both factors related to the photon source demultiplexer, $RR=80$~MHz, the repetition rate, and $\eta_{1}$, the single-channel transmittance.} is $R_{tot} (\propto \eta_{1}^{5}) \approx 92$~nHz--$47$~mHz.
\new{In five years, systems with 2~GHz clock rates and total efficiency of $\eta_{tot} {>} 90\%$ are projected to be available, leading to five-photon rates of ${\sim}1.2$~GHz, an improvement of sixteen to eleven orders-of-magnitude.} \new{At that point QPU-driven quantum-accelerated machine learning offers capabilities orders of magnitude greater than those possible with conventional compute for the quantum fingerprint}~\footnote{\new{
For example, our computation of the quantum fingerprint for the full 70,000-image MNIST dataset using 5 photons in 24 modes for 10 unitaries takes around 2,800 CPU-hours and 6.5~TB of RAM. 
On a physical QPU with a 5-photon rate of ${\sim}1.2$~GHz, resolving all D{=}42,504 outcomes to comparable statistical precision requires $10^6*70,000*10 = 7*10^{11}$ (samples $*$ images $*$ unitaries) five-fold detection events, corresponding to around 10 minutes of acquisition time.}}.

This work presents a comprehensive analysis of quantum optical reservoir computing, demonstrating its capability for practical quantum acceleration of classification tasks.
QORC consistently outperforms classical linear classifiers across a range of tasks, even for the limits of completely distinguishable photons or a single photon input, $N{=}1$, where only first-order quantum coherence occurs.
The scheme achieves comparable or superior test accuracies while requiring around twenty times fewer training images, highlighting the data-efficiency advantage and acceleration.
Crucially, the advantages are experimentally verified on a physical QPU, the Ascella processor by Quandela.
The very close quantitative and qualitative agreement between numerical results and experiments on a real quantum photonic processor confirms the physical validity of our reservoir approach.
Moreover, QORC shows strong improvements on imbalanced and complex datasets, including artificially imbalanced MNIST and medical images from MedMNISTv2.

Future research directions will include: expanding QORC's advantage on physical hardware, focusing on runtime, scalability, and efficiency;
and new techniques that preserve the unique structural fingerprint while reducing the dimensionality of the distribution~\cite{seron_efficient_2024,anguita_experimental_2025}, reducing the inputs to the classifier and hence further accelerating the classification.
\new{Extending our framework beyond static toy model image data to e.g. complex datasets like ImageNet~\cite{Deng2009} or pattern recognition in time-series~\cite{kobayashi_feedback-driven_2024,jafarigol_aiml-based_2025} will help to identify potential overfitting risks and generalisation failure of the scheme and could reveal new regimes of quantum reservoir performance, particularly for dynamical feature extraction.}
Ultimately, advancing both the hardware implementation and task diversity of QORC will be key to realising concrete quantum advantage in real-world machine learning tasks.

\vspace{2mm}
\noindent \emph{Acknowledgements} We thank Michael Harvey, Jean Senellart, and Sally Shrapnel for valuable discussions.
MR is the recipient of a Fellowship funded by the Big Questions Institute.
KN, WJM and AS were partially supported by the MEXT Quantum Leap Flagship Program (MEXT Q-LEAP) under Grant No. JPMXS0118069605, the Japanese Cross-ministerial Strategic Innovation Promotion Program (SIP) under Grant No. JPJ012367, COI-NEXT (JPMJPF2221) and the JSPS Kakenhi Grant No. 21H04880.
AGW is the recipient of an Australian Research Council Laureate Fellowship (FL210100045) funded by the Australian Government. 
This research was supported in part by the Australian Research Council Centre of Excellence for Engineered Quantum Systems (EQUS, CE170100009).
\new{This work was supported by resources provided by The University of Queensland Research Computing Centre’s Bunya supercomputer~\cite{DellTechnologies2024}, with funding from The University of Queensland, Brisbane, Australia.}

\vspace{2mm}
\noindent \emph{Data availability} The numerical and experimental data used in this paper will be made publicly available in a repository upon publication.

\vspace{1mm}
\noindent \emph{Code availability} Custom code used for data pre-processing, hybrid classical-quantum input encoding, and numerical simulations and QPU experiments was developed for this study.
Representative scripts sufficient to reproduce the main results are available from the corresponding author upon reasonable request.

\bibliographystyle{myapsrev4-2}
\bibliography{references}

\end{document}


\title{Supplemental Material: Photonic Quantum-Accelerated Machine Learning}

\author{Markus Rambach}
\email{m.rambach@uq.edu.au}
\affiliation{Quantum Technology Laboratory, School of Mathematics and Physics, The University of Queensland, Brisbane, QLD 4072, Australia}

\author{Abhishek Roy}
\affiliation{Quantum Technology Laboratory, School of Mathematics and Physics, The University of Queensland, Brisbane, QLD 4072, Australia}
\affiliation{Department of Physics and Astronomy, Macquarie University, Sydney, NSW 2113, Australia}

\author{Alexei Gilchrist}
\affiliation{Department of Physics and Astronomy, Macquarie University, Sydney, NSW 2113, Australia}

\author{Akitada Sakurai}
\affiliation{Okinawa Institute of Science and Technology Graduate University, Onna-son, Okinawa 904-0495, Japan}

\author{William J. Munro}
\affiliation{Okinawa Institute of Science and Technology Graduate University, Onna-son, Okinawa 904-0495, Japan}

\author{Kae Nemoto}
\affiliation{Okinawa Institute of Science and Technology Graduate University, Onna-son, Okinawa 904-0495, Japan}

\author{Andrew G. White}
\affiliation{Quantum Technology Laboratory, School of Mathematics and Physics, The University of Queensland, Brisbane, QLD 4072, Australia}

\maketitle

\section{DIGITS data set}
\label{sec:digits}
The DIGITS dataset~\cite{e_alpaydin_pen-based_1996} is a classic benchmark in machine learning, consisting of 8-by-8 greyscale images of handwritten digits (0-9).
Each image contains pixel intensity ranging from 0 to 16 in a compact 64-feature representation.
The dataset includes 1797 samples, making it lightweight and well-suited for testing algorithms and evaluating classification performance on simple visual tasks and suitable for the current capabilities of a QPU.
Compared to datasets like MNIST or MedMNISTv2, DIGITS is smaller and lower resolution, which allows for fast experiments while still capturing key challenges of image recognition like shape variation and noise.

\begin{figure}[!b]
\centering
\vspace{3mm}
\includegraphics[width=\linewidth]{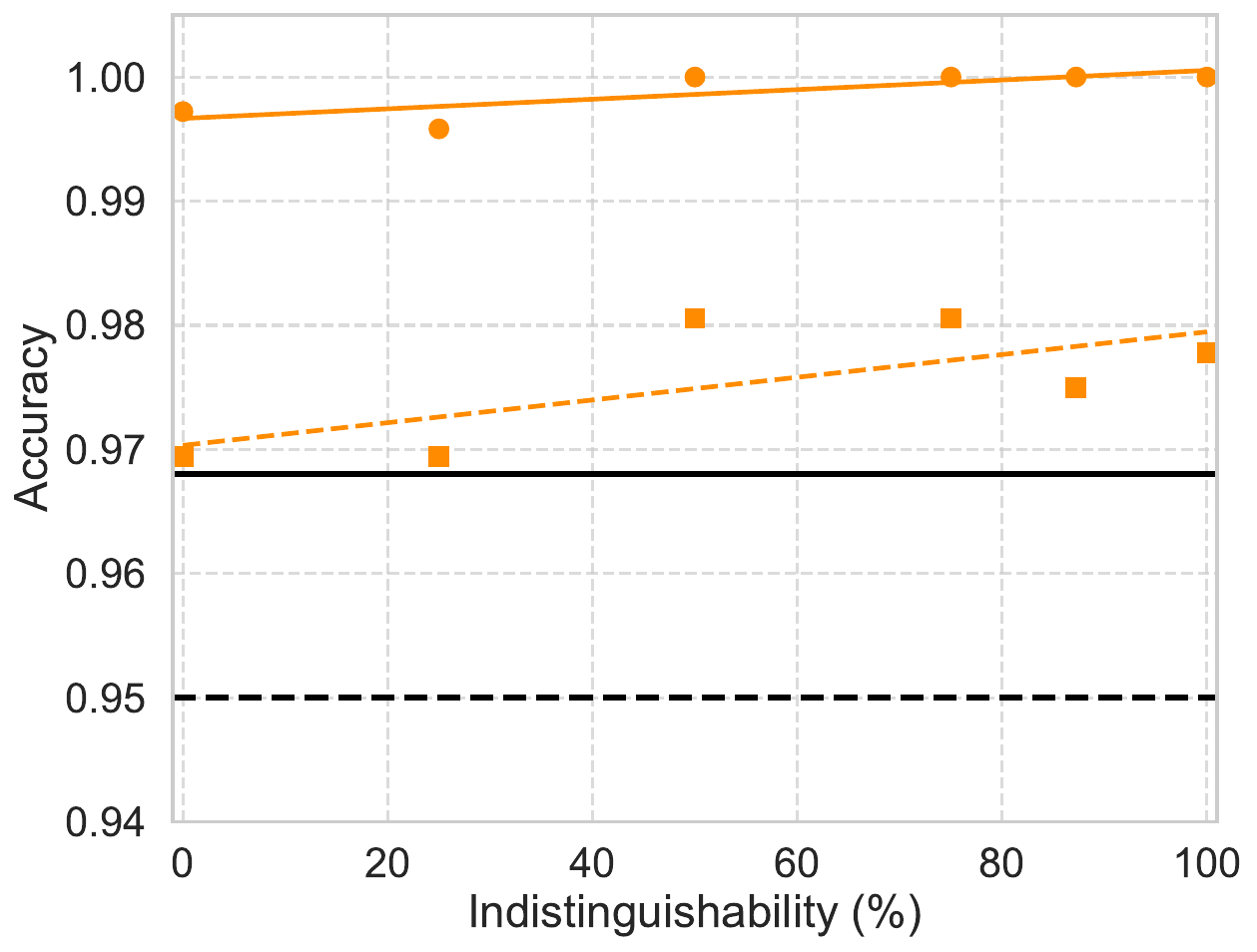}
\caption{\label{fig:DIGITS_HOM312}
\justifying
DIGITS classification accuracy as a function of single photon indistinguishability.
QORC with $N{=}3, M{=}12$ in orange, MLR in black for comparison.
Training as circles and solid lines for 100 epochs, testing as squares and dashed lines.
Train/test split is 80/20 (1437/360 images).
For each image, the reservoir is 30,000
samples, noise encoded via non-ideal $g_2$ and indistinguishability.
Orange fitted lines to guide the eye only.
All accuracies for all indistinguishabilities surpass the maximal achievable training accuracy of MLR.
}
\end{figure}

Fig.~\ref{fig:DIGITS_HOM312} shows the dependence of the classification accuracy of QORC on the photon indistinguishability for $N{=}3, M{=}12$ (orange) and compares it to the MLR (black).
The results are noisier than in the MNIST case, with some lower indistinguishabilities performing better than higher ones, however, we believe this is a shot-to-shot effect and will change with other unitaries or a different train/test split.
The improvements range from 1.9\% to 3.1\% for the test accuracies, and 2.8\% to 3.2\% on training data.
The training accuracy is 100\% for all indistinguishabilities above 25\%.

\section{Confusion Matrix}
\label{sec:confusion}

\begin{figure*}[t]
\centering
\includegraphics[width=\linewidth]{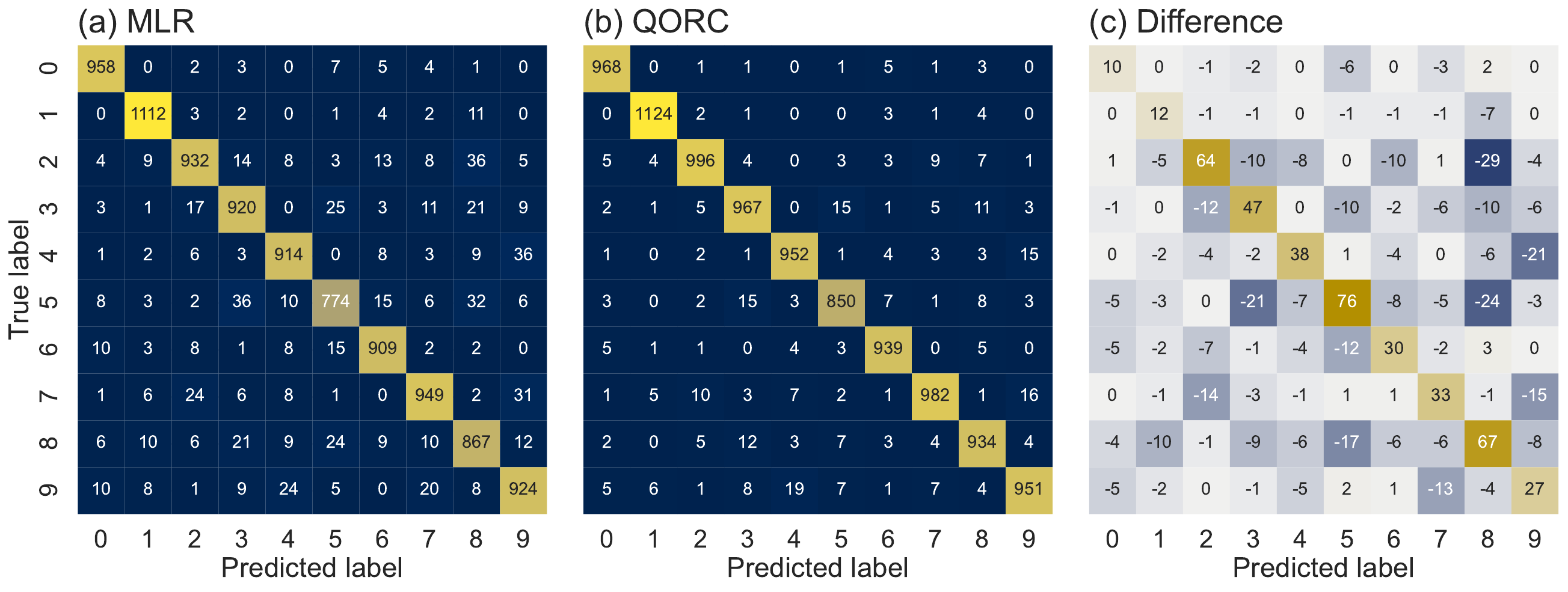}
\caption{\label{fig:MNIST_confusion}
\justifying
Confusion matrix examples for MNIST classification on test data, numbers in the matrices indicate classification instances.
Training on 60,000 images for 100 epochs, testing on 10,000 images.
Reservoir: 30,000 samples.
(a) Linear classifier (MLR), (b) QORC, (c) Difference: QORC minus MLR.
Difference is plotted with a different colour map for clarity: shades of gold for positive values, light grey for no difference (values around 0), and shades of dark blue for negative values.}
\end{figure*}

A confusion matrix is the visualisation of a table that summarises the performance of a classification model by comparing predicted and true class labels.
It shows how often each class is correctly identified (diagonals) and where misclassifications happen (off-diagonals), providing detailed insight beyond overall accuracy, which is the ratio of the two.
Fig.~\ref{fig:MNIST_confusion} a), b) illustrates examples of such matrices for the MLR and QORC, respectively.
They look very similar as the achieved accuracies---defined as the sum of the diagonal over the sum of all values---are above 90\% in both cases and differences are subtle.
These differences are highlighted individually in Fig.~\ref{fig:MNIST_confusion} c) which shows a similar confusion matrix representation for QORC b) minus MLR a).
We clearly see that all values on the diagonal are positive, showing the increase in correctly classified images.
Similar, all off-diagonal elements are ${\leq}0$, showing a decrease in incorrect classification events.
The seemingly larger differences of correct classifications between different classes, e.g., \textit{zero} vs. \textit{one}, mainly stem from slight class imbalances in the test dataset.

\section{Reproducibility}
\label{sec:reproducibility}
Reproducibility in training classifiers ensures that results can be reliably replicated under the same conditions, which is essential for scientific validity and trust in the model.
It allows for the verification of findings, fair comparison of methods, and build upon prior work with confidence.
We investigate the differences in accuracies when using different unitaries and/or input states to the boson sampling network ($N{=}3, M{=}12$) that creates our quantum data for acceleration.
Table~\ref{tab:MNIST_MC} lists the mean and standard deviation of the accuracy of QORC with 10 independent runs of boson sampling, all collecting 30,000 samples per image.
We can see that the results are very similar and independent of the applied random unitary or the selected input channels of the QPU.
The standard deviation is small compared to the achieved advantage of QORC over MLR and the means lie well within 1$\sigma$.

\begin{table}[h]
\caption{\label{tab:MNIST_MC}
\justifying
MNIST accuracy on Monte Carlo simulations, 100 training epochs.
Reservoir: 30,000 samples, $N{=}3, M{=}12$.
10 independent sampling experiments, changing only the value for the \textit{Parameter} column.
STD: standard deviation.}
\begin{tabular}{lcccc}
\toprule
& \multicolumn{2}{c}{Train Accuracy (\%)} & \multicolumn{2}{c}{Test Accuracy (\%)} \\
\cmidrule(lr){2-3}
\cmidrule(lr){4-5}
Parameter & {Average} & {STD} & {Average} & {STD} \\
\midrule
Reservoir unitary     & 96.87 & 0.18 & 95.62 & 0.18 \\
Input state & 96.81 & 0.13 & 95.57 & 0.13 \\
\bottomrule
\end{tabular}
\end{table}

\section{Increasing photon number}
\label{sec:M20}

In Table~\ref{tab:MNIST_M20} we analyse the dependence of the achievable accuracy on the photon number $N$ for a fixed number of modes $M{=}20$.
Even with just a single photon in the circuit, we see small but measurable improvements over the case of just inputting the 20 PCA components as additional inputs. This reflects that even a single photon can interfere with itself in the circuit, producing a nonlinear response that adds new information to the neural network.
We also see that the accuracy advantage gets smaller with every additional photon. Note that the boson sampling output space grows combinatorially, so we've increased the number of samples for accurate sampling.

\begin{table}[h]
\caption{\label{tab:MNIST_M20}
\justifying
MNIST accuracies for different photon inputs.
Reservoir: $N$ photons, $M{=}20$ modes, 30,000 samples (100k for $N{=}4$, 1M for $N{=}5$).
Training for 100 epochs.
Added Inputs is the number of additional inputs, i.e., all possible distributions of $N$ photons into $20$ modes.
$\Delta$MLR is the improvement of test accuracy in QORC over MLR.
}
\begin{tabular}{lcccc}
\toprule
Photons & Added Inputs & Train Acc. (\%) & Test Acc. (\%) & {$\Delta$MLR}\\
\midrule
1 & 20   & 94.2  & 93.2 & 0.6 \\
2 & 190  & 96.6  & 95.2 & 2.6 \\
3 & 1,140 & 98.9  & 96.5 & 3.9 \\
4 & 4,845 & 100.0 & 97.4 & 4.8 \\
5 & 15,504& 100.0 & 97.5 & 4.9 \\
\bottomrule
\end{tabular}
\end{table}

\begin{figure}[h!]
\centering
\begin{overpic}[width=\columnwidth]{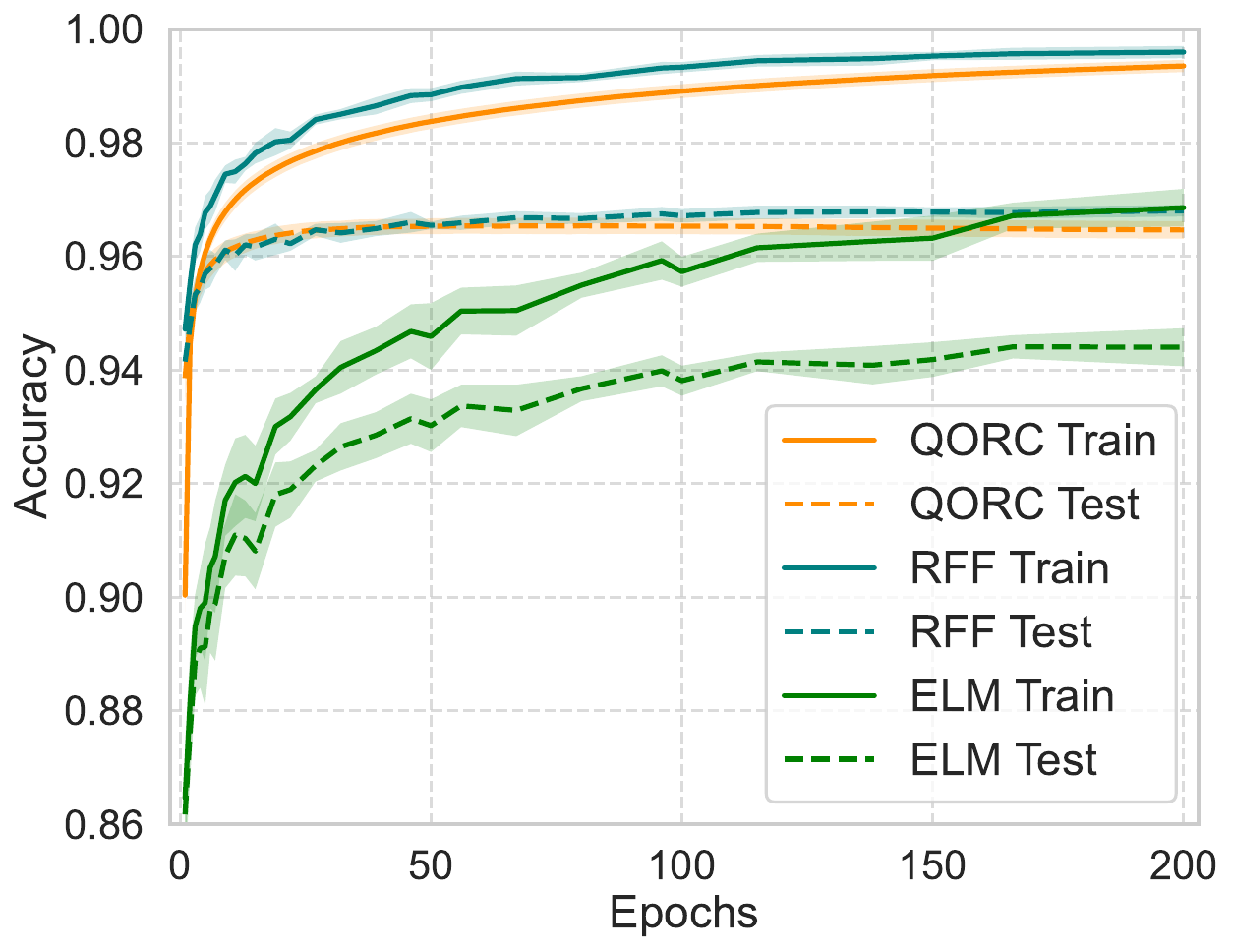}
    \put(2,80){\large\textbf{(a)}}
\end{overpic}
\hfill
\vspace{2mm}
\begin{overpic}[width=\columnwidth]{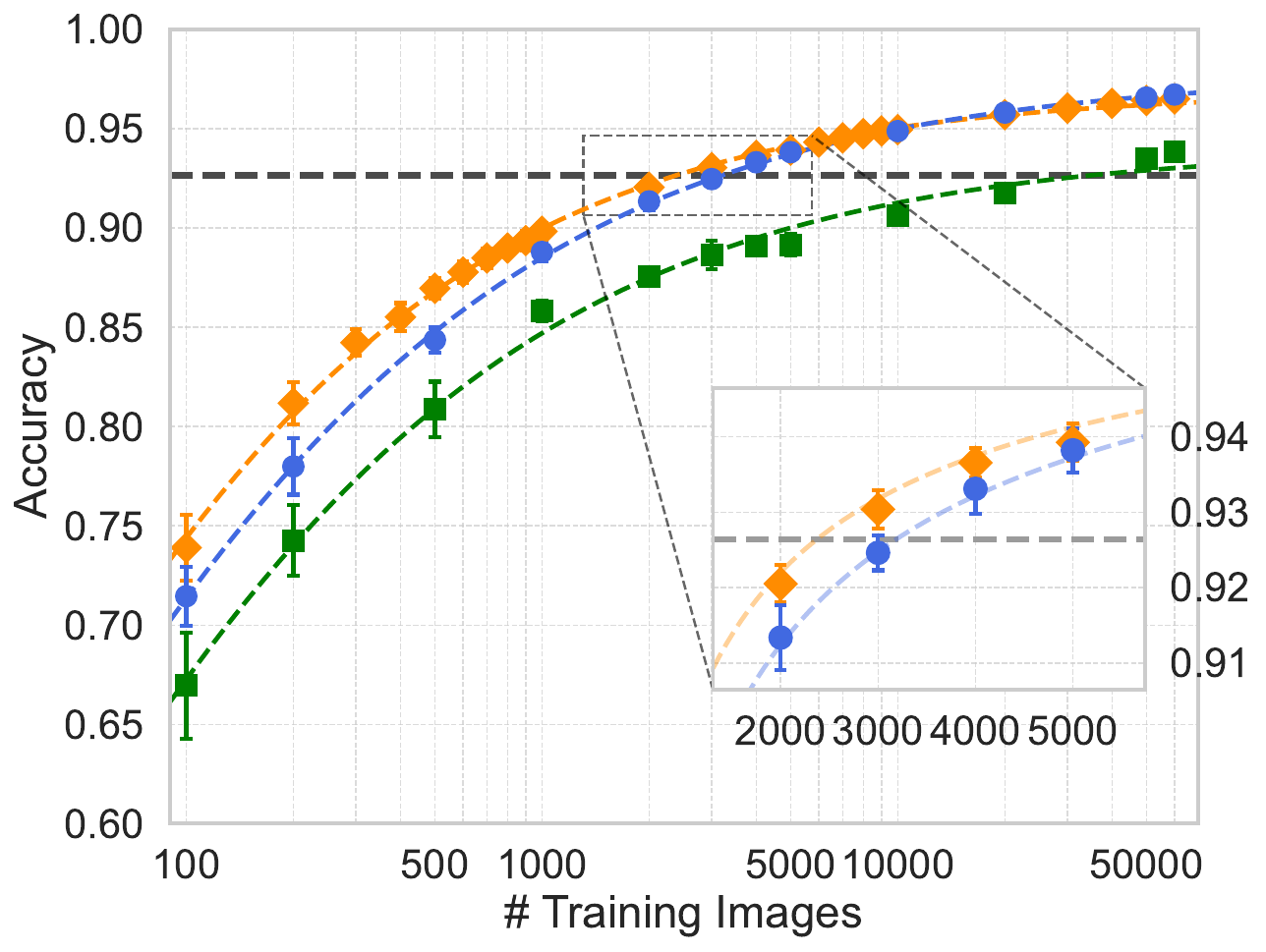}
    \put(2,80){\large\textbf{(b)}}
\end{overpic}
\caption{\justifying
    \new{QORC vs. RFF vs. ELM.
    QORC in orange ($N{=}3, M{=}20$), RFF in blue, ELM in green (both with $D{=}1,140$).
    Training (solid lines) on 60,000 images, testing (dashed lines) on 10,000 images.
    For each image, the reservoir is 30,000 samples.
    Shaded region/error bars are the mean of 100 classification runs: 10 different unitaries with 10 different train/test splittings.
    (a) Accuracy as a function of training epochs.
    QORC and RFF perform very similar, both outperform ELM.
    Best test performance for QORC/RFF/ELM is reached around 40/100/200 epochs.
    Shaded region for QORC and RFF is very small.
    (b) MNIST classification test accuracy dependent on (balanced) training dataset size $n_{tr}$. 
    QORC (noiseless) with $N{=}3, M{=}20$ in orange diamonds, RFF in blue dots, ELM in green squares.
    Grey dashed line is the best possible performances for test accuracies with MLR.
    Training for 100 epochs, testing (dashed lines) on $n_{te}{=}$10,000 images.
    Dashed coloured lines to guide the eye only, Hill function fit with function $L / (1 + (x_{50} / x)^k)$.
    }}
\label{fig:RFF_ELM}
\end{figure}

{
\color{black}
\section{Classical analogues: random Fourier features and extreme learning machines}
\label{sec:RFF_ELM}

Both baselines considered here share the defining structure of a reservoir computer: a \emph{fixed}, untrained random nonlinearity that lifts the input into a high-dimensional feature space, followed by a linear readout that is the only trained component. 
Random Fourier features (RFF)~\cite{Rahimi2007} realise this with sinusoids of randomly drawn frequencies, approximating a shift-invariant kernel; 
extreme learning machines (ELM)~\cite{Huang2006} realise it with a single hidden layer of randomly initialised weights and a point-wise nonlinearity. 
Neither involves a quantum resource, and in neither case is the random layer optimised. 
To keep the comparison strict, the readout, training regime and evaluation protocol are identical to those used for QORC in Table~I: 
multinomial logistic regression (MLR) trained for $100$ epochs on the same splits, with the same accuracy and $\Delta$MLR definitions.

\subsection{Random Fourier features}
\label{sec:rff_setup}

The number of random features is matched to the number of additional inputs supplied to the MLR by QORC at each configuration, so that the two models present the classifier with vectors of equal length: $D=220$ for $(N,M)=(3,12)$, $D{=}1,140$ for $(3,20)$, and $D{=}42,504$ for $(5,24)$. 
Features are generated in the standard cosine form
\begin{equation}
  \phi_j(\mathbf{x}) \;=\; \sqrt{\tfrac{2}{D}}\,
  \cos\!\left(\boldsymbol{\omega}_j^{\!\top}\mathbf{x} + b_j\right),
  \qquad j = 1,\dots,D ,
  \label{eq:rff}
\end{equation}
with frequencies $\boldsymbol{\omega}_j$ and biases $b_j$ drawn independent and identically distributed, 
which yields an unbiased estimate of the Gaussian (RBF) kernel $k(\mathbf{x},\mathbf{x}') = \exp(-\lVert \mathbf{x}-\mathbf{x}'\rVert^2/2\sigma^2)$.
The bandwidth was set from the variance of the quantum fingerprints on the training set and then rescaled by factors $\{0.5,\,1.0,\,2.0\}$ to verify that the comparison is not an artefact of a single kernel width. 
We see no dependence on the variance and the values quoted in Table~\ref{tab:s8} use the $1.0\times$ setting. 
Once drawn, $\{\boldsymbol{\omega}_j, b_j\}$ are frozen for the entire run.

\subsection{Extreme learning machines}
\label{sec:elm_setup}

The ELM hidden layer is sized identically ($D = 220$, 1,140, 42,504 for the three configurations), so RFF, ELM and QORC all hand the readout a feature vector of the same dimension. 
Hidden activations are $\mathbf{h}(\mathbf{x}) = g(W\mathbf{x} + \mathbf{b})$ with $g$ the sigmoid nonlinearity, $W_{ij} \sim \mathcal{N}(0,\sigma_w^2)$ and $\mathbf{b} \sim \mathcal{U}[-1,1]$.
As for RFF, $\sigma_w^2$ was scanned over $\{0.5,\,1.0,\,2.0\}$ times its nominal value. 
We emphasise that $W$ and $\mathbf{b}$ are sampled once and never
updated: gradients are applied to the MLR readout alone, over the same $100$ epochs used throughout, so that ELM differs from RFF only in the form of the random nonlinearity and not in what is trained.

All entries in Table~\ref{tab:s8} are means over $100$ independent random draws of the feature map; the quoted uncertainties are standard deviations across those draws. 
Figure~\ref{fig:RFF_ELM} shows the corresponding train-size and
epoch scans at $(N,M)=(3,20)$.

\subsection{Optimisation cost}
\label{sec:scaling}

An exact RBF kernel machine must assemble and factorise the $n \times n$ Gram matrix, costing $\mathcal{O}(n^3)$ time and $\mathcal{O}(n^2)$ memory in the number of training images. 
QORC incurs no such cost: the reservoir maps each image independently of every other, and the MLR readout is optimised by first-order updates costing $\mathcal{O}(nD)$ per epoch, i.e.\ linear in $n$ at fixed $D$.

RFF and ELM were designed for precisely this purpose, however---replacing the implicit kernel by an explicit $D$-dimensional map so that training reduces to a linear problem---and under the identical readout they are likewise $\mathcal{O}(nD)$ per epoch. 
All three models of Table~\ref{tab:s8} therefore share the same linear scaling in $n$; only the exact kernel machine pays $\mathcal{O}(n^3)$.

What separates them is the cost of producing the map. 
RFF and ELM require $\mathcal{O}(D)$ arithmetic per image, with $D{=}42,504$ already at $(N,M)=(5,24)$ and growing combinatorially thereafter, whereas the interferometer uses a physical process in a single optical pass. 
QORC's throughput is thus set by photon-source brightness and detection efficiency rather than by available computational power, which is the sense in which the advantage reported in the main text is one of acceleration. 

\begin{table}[t]
  \caption{
  \justifying
  \new{
  Classical analogues compared with QORC under the identical MLR
  training regime ($100$ epochs). $D$ is the feature-space dimension presented
  to the readout, matched across models at each $(N,M)$. Values are means over
  $100$ random feature draws.
  }}
  \label{tab:s8}
  \begin{ruledtabular}
  \begin{tabular}{llrcc}
    $(N,M)$ & Model & $D$ & Train (\%) & Test (\%)  \\
    \colrule
    \multirow{3}{*}{$(3,12)$}
      & QORC & 220             & 96.6 (1) & 95.6 (1)  \\
      & RFF  & 220             & 96.8 (3) & 94.3 (3)  \\
      & ELM  & 220             & 93.7 (4) & 91.7 (4)  \\
    \colrule
    \multirow{3}{*}{$(3,20)$}
      & QORC & 1,140      & 98.9 (1) & 96.5 (1)  \\
      & RFF  & 1,140      & 99.3 (1) & 96.7 (1)  \\
      & ELM  & 1,140      & 95.7 (3) & 93.8 (3)  \\
    \colrule
    \multirow{3}{*}{$(5,24)$}
      & QORC & 42,504     & 100.0 (0) & 97.5 (1)  \\
      & RFF  & 42,504     & 99.7 (1)  & 97.5 (1)  \\
      & ELM  & 42,504     & 96.8 (4)  & 95.2 (4)  \\
  \end{tabular}
  \end{ruledtabular}
\end{table}

\subsection{Discussion}
\label{sec:rff_elm_discussion} 

The pattern of Table~\ref{tab:s8} is reproduced in Fig.~\ref{fig:RFF_ELM}.
Panel~(a) exhibits close agreement between the QORC and RFF curves across the full range, whereas the ELM curve lies systematically below both and displays markedly larger variance. 
Panel~(b) recovers the same ordering for the training acceleration, though the separation is more pronounced in this metric. Note that QORC attains a speedup of $25.3\times$, compared with $18.8\times$ for RFF and $1.7\times$ for ELM. 
Since RFF reproduces the QORC accuracy to within error, the two schemes are not distinguished by the content of their feature maps, but we can see a difference by the initial rate at which our scheme learns, especially in the resource-sparse regime. 

RFF performs similarly to QORC in most cases because the two are structurally similar objects: a fixed random projection into a high-dimensional space, followed by a trained linear separator.
The sinusoidal map of Eq.~\eqref{eq:rff} and the Fock-space map realised by the interferometer differ in the induced kernel but not in kind. 
ELM trails RFF at equal $D$, indicating that a random dense layer with a point-wise nonlinearity distributes its capacity less efficiently over the same number of coordinates than either the sinusoidal or the quantum fingerprint.

The advantage of QORC over an exact RBF kernel machine is the avoidance of $\mathcal{O}(n^3)$ optimisation, a property it shares with RFF and ELM, while its advantage over RFF and ELM lies in how the high-dimensional map is produced rather than in what the map contains or in how the readout is trained.
This and the improved training data efficiency are additional acceleration arguments to the main text, with the results in this section establishing its scope. 
}

\section{Imbalanced datasets}
\label{sec:imbalanced}

\subsection{Macro F1 score}
The macro F1 score is a balanced measure of classifier performance across multiple classes, giving equal weight to each class regardless of its frequency.
It is defined as the arithmetic mean of the per-class scores F$_{1, i}$ defined as
\begin{equation}
    \text{F}_{1, i}=\frac{2 P_i R_i}{P_i+R_i},
\end{equation}
where $P_i$ is the precision and $R_i$ is the recall of sub-class $i$.
Precision and recall are defined as
\begin{equation}
P=\frac{TP}{TP+FP}, \quad R=\frac{TP}{TP+FN},
\end{equation}
with $TP$, $FP$, and $FN$ the true positive, false positive, and false negative of the confusion matrix.

For example, let's consider the digit \textit{7}: a true positive is when the true label and the prediction is \textit{7}.
A false positive is when the prediction is still \textit{7}, but the true label is anything but the prediction.
Finally, for a false negative the true label is a \textit{7}, but the prediction is any other category.
The macro F1 score is then given as
\begin{equation}
    \text{F}_1^{\text {macro }}=\frac{1}{C} \sum_{i=1}^C F_{1, i}~,
\end{equation}
mitigating all potential imbalances in training and test datasets.

\subsection{Datasets}

\subsubsection{MNIST}
Table~\ref{tab:weightsMNIST} shows training set weights for differently balanced artificial datasets on MNIST.
In the Gaussian case, the majority class in training is \textit{five}.
It accounts for $1/5$ of the total training images and includes around $17{\times}$ more training images than the smallest minority class \textit{zero}.
For the severely imbalanced dataset, which was modelled loosely after the medical DermaMNIST dataset below, the majority class is over $100{\times}$ more prominent than the smallest minority class, and makes up for ${\sim}2/3$ of all training images.
To remove potential biases, we randomly assigned the class weights to actual digits in each run for the severely imbalanced case and investigated 100 different training runs.

As discussed in the main text, Table~\ref{tab:MNIST_balanceF1} shows that QORC improves the F1 score for all datasets on both training and test images, demonstrating its significant value for real-world applications.
Although not as meaningful as the F1 score, Table~\ref{tab:MNIST_balance} shows all the achieved accuracy values on the artificially imbalanced MNIST data sets.
We can see that the advantage of QORC increases for larger imbalance; however, the overall test accuracy is significantly lower (as expected).

\begin{table}[h]
\caption{\label{tab:weightsMNIST}
\justifying
MNIST imbalance.
For ``Gauss'': $\mu{=}4.75, \sigma{=}2$ to keep it centred but slightly asymmetric.
For ``Imbal'': the percentages are randomly shuffled onto the categories in each run.
See main text for further details.}
\begin{tabular}{l|cccccccccc}
\toprule
Data & \multicolumn{10}{c}{Class Percentages (Numbers 0-9)}\\
\midrule
Bal   & 10 & 10 & 10 & 10 & 10 & 10 & 10 & 10 & 10 & 10 \\
Gauss & 1.2 & 3.5 & 7.9 & 13.8 & 18.8 & 20.0 & 16.6 & 10.7 & 5.4 & 2.1 \\
Imbal & 64.3 & 9.7 & 8.0 & 6.4 & 3.9 & 3.2 & 1.6 &  1.3 &  1.0 & 0.6 \\
\bottomrule
\end{tabular}
\end{table}

\begin{table}[h]
\caption{\label{tab:MNIST_balanceF1}
\justifying
MNIST classification macro F1 scores for differently balanced datasets, QORC for $N{=}3, M{=}20$.
Values are the mean of 100 different subsets of training data, errors (in brackets) are the standard deviation.
Training for 100 epochs with ${\sim}$10,000 train images, evaluation on 10,000 test images.
Top: balanced training data between all categories;
Middle: Gaussian filter with $\mu{=}4.75, \sigma{=}2$;
Bottom: Severely imbalanced training data with ${\sim}2{\times} (64.3\%)$ images in majority category compared to all other categories.}
\begin{tabular}{lcc}
\toprule
&  \multicolumn{2}{c}{Macro F1 Score}  \\
\cmidrule(lr){2-3}
Input Data & {Train} & {Test}  \\
\midrule
Balanced (MLR) & 0.941 (2)  & 0.914 (2)  \\
Balanced (QORC)  & 0.9998 (7) & 0.949 (2)  \\
\midrule
Gaussian (MLR) & 0.993 (4) & 0.886 (3)  \\
Gaussian (QORC)  & 0.9998 (2) & 0.931 (3)  \\
\midrule
Severely Imbalanced (MLR) & 0.929 (9) & 0.846 (12) \\
Severely Imbalanced (QORC)  & 0.9999 (1) & 0.894 (8)  \\
\bottomrule
\end{tabular}
\end{table}

\begin{table}[h]
\centering
\caption{\label{tab:MNIST_balance}
\justifying
MNIST classification accuracies for differently balanced datasets, QORC for $N{=}3$, $M{=}12/20$.
Training for 100 epochs with ${\sim}$10,000 train images, evaluation on 10,000 test images.
Top: balanced training data between all categories;
Middle: Gaussian filter with $\mu{=}4.75, \sigma{=}2$;
Bottom: Severely imbalanced training data with ${\sim}2{\times} (64.3\%)$ images in majority category compared to all other categories.
$\Delta$MLR is the difference to MLR.}
\begin{tabular}{lcccc}
\toprule
& & \multicolumn{3}{c}{Accuracy (\%)}  \\
\cmidrule(lr){3-5}
Input Data & Modes & {Train} & {Test} & {$\Delta$MLR} \\
\midrule
Balanced & {\text{--}} & 93.6  & 91.4 & {\text{--}} \\
Balanced & 12          & 98.8  & 94.4 & 3.0 \\
Balanced & 20          & 100.0 & 95.0 & 3.6 \\
\midrule
Gaussian & {\text{--}} & 95.5  & 89.1 & {\text{--}} \\
Gaussian & 12          & 99.1  & 92.4 & 3.3 \\
Gaussian & 20          & 100.0 & 93.6 & 4.5 \\
\midrule
Severely Imbalanced & {\text{--}} & 96.6 & 84.7 & {\text{--}} \\
Severely Imbalanced & 12          & 99.6 & 89.0 & 4.3 \\
Severely Imbalanced & 20          & 100.0 & 90.0 & 5.3 \\
\bottomrule
\end{tabular}
\end{table}

\subsubsection{MedMNISTv2}
\begin{figure}[!t]
\centering
\includegraphics[width=\linewidth]{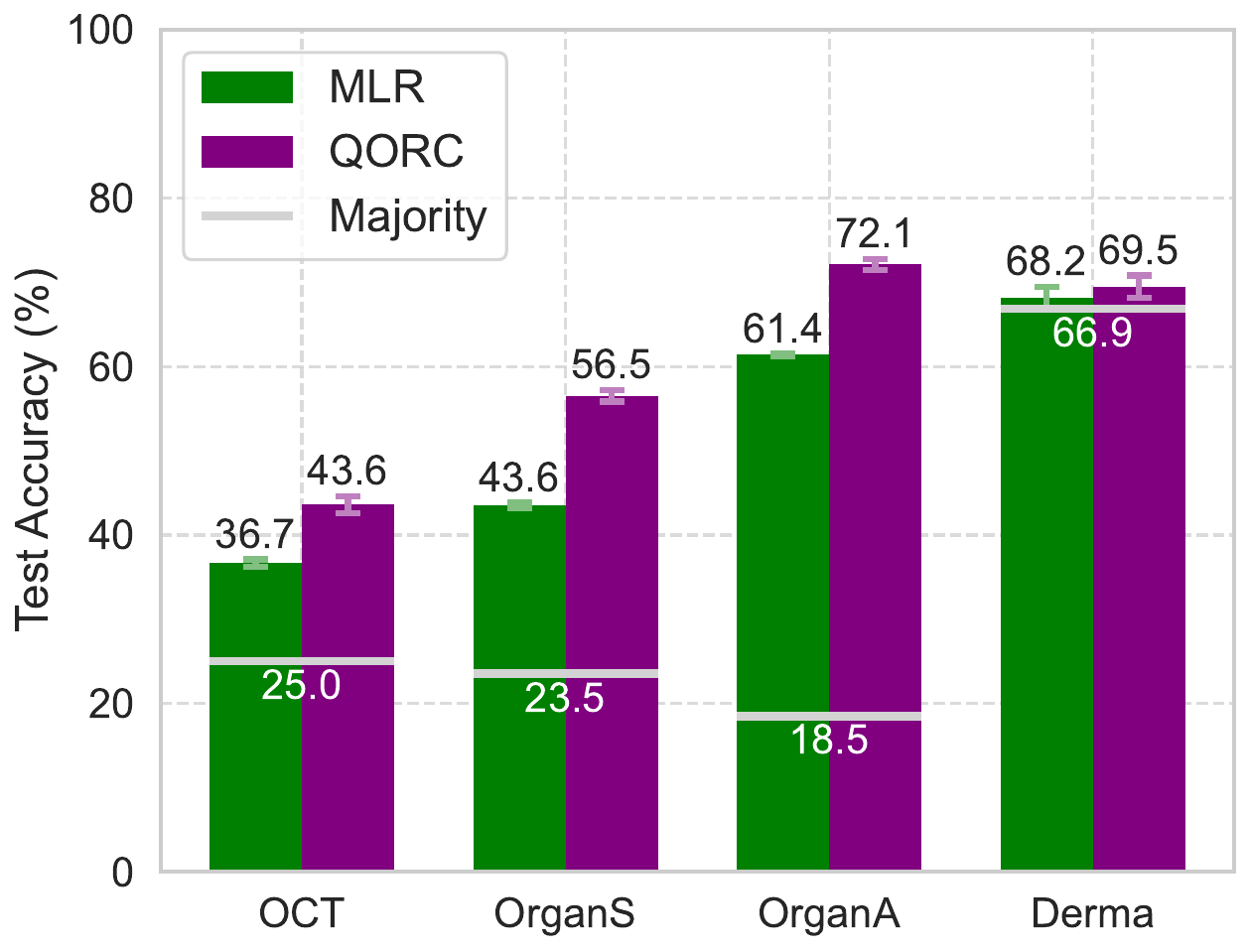}
\caption{\label{fig:MedMNIST}
\justifying
QORC classification accuracy improvements on different datasets in MedMNISTv2 for 200 training epochs.
\new{Errors bars are the mean of 100 classification runs: 10 different unitaries with 10 different train/test splittings.}
Quantum reservoir: $N{=}3, M{=}20$; for each image, the reservoir is 30,000
samples.
Majority: classifier which just picks the majority training class each time without learning, \new{then assigns that label to all images in the test set.}
}
\end{figure}

Table~\ref{tab:weightsMed} presents the training set imbalances for the investigated datasets in MedMNISTv2.
Additionally, it highlights the maximally achievable test accuracies with deep neural networks~\cite{yang_medmnist_2023}, putting the results from Fig.~\ref{fig:MedMNIST} into perspective.
Please see Ref.~\cite{yang_medmnist_2023} for further details and metrics.

Fig.~\ref{fig:MedMNIST} illustrates achieved accuracies on MedMNISTv2.
We observe that the overall performance of MLR and QORC is significantly worse than for imbalanced MNIST, indicating that the algorithms struggle mainly with the higher complexity of the images rather than the imbalanced training data.
For comparison, we also added the majority classification, where we just pick the majority class and assign its label to all test images.
This leads to almost identical results on accuracy for the MLR on the OCT and especially Derma datasets, showing that little is actually being learned by the classical model in both cases.
Although the maximally achievable accuracies of deep neural networks are outside the reach of QORC (although not too far away for OrganS and Derma), our scheme again outperforms MLR in all metrics on all investigated datasets.

\section{KerasTuner details}
\label{sec:KerasTuner}
Table~\ref{tab:KTpar} presents the optimised parameter space and Table~\ref{tab:KTval} summarises the optimal values found by KerasTuner.

\begin{table*}[h]
\centering
\caption{\label{tab:weightsMed}
\justifying
MedMNISTv2 class imbalances and maximally achievable test accuracies on deep neural networks according to Ref.~\cite{yang_medmnist_2023}.
OCT (optical coherence tomography images for retinal disease, greyscale),
Organ{A,S} ({Axial, Sagittal} CT images from liver tumour segmentation, greyscale),
and Derma (dermatoscopic images of pigmented skin lesions, colour)}
\begin{tabular}{l|ccccccccccc|c}
\toprule
Data  &  \multicolumn{11}{c}{Class Percentages} & Acc. (\%)\\
\midrule
OCT     & 34.4 & 10.5 & 8.0  & 47.2 & -- & -- & -- & -- & -- & -- & -- & 77.6\\
OrganS  & 8.2  & 4.5  & 4.4  & 5.2  & 8.1  & 8.0  & 24.9 &  5.3 &  5.8 & 14.4 & 11.2 & 81.3 \\
OrganA  & 5.7  & 4.0  & 3.9  & 4.3  & 11.5 & 11.0 & 17.8 & 11.3 & 11.4 & 8.8  & 10.3 & 95.1 \\
Derma   & 3.3  & 5.1  & 11.0 & 1.1  & 11.1 & 67.0 & 1.4  & -- & -- & -- & -- & 76.8 \\
\bottomrule
\end{tabular}
\end{table*}

\begin{table*}[h]
\caption{\label{tab:KTpar}
Optimisation space for different model architectures.
\cmark~indicates parameters that are optimised, \xmark~parameters that are either not optimised due to time constrains or not applicable (e.g. hidden layers in a linear network).
}
\begin{tabular}{l|ccccccc}
\toprule
& \multicolumn{7}{c}{Parameters} \\
\cmidrule(lr){2-8}
Model & Hidden Layers & Activation Function & Units per Layer & Dropout Rate & Optimiser & Learning Rate & Batch Size \\
\midrule
Linear   & \xmark & \xmark & \xmark & \xmark & \cmark & \cmark & \cmark   \\
Shallow (fixed) & \xmark & \cmark & \xmark & \cmark & \cmark & \cmark & \cmark    \\
Shallow & \xmark & \cmark & \cmark & \cmark & \cmark & \cmark & \xmark  \\
Deep & \cmark & \cmark & \cmark & \cmark & \cmark & \cmark & \xmark  \\
\bottomrule
\end{tabular}
\end{table*}

\begin{table*}[h]
\centering
\caption{\label{tab:KTval}
\justifying
Optimised parameter values for different model architectures: Linear: MLR; ShallowF: one hidden layer with fixed number of units (16); Shallow: same as ShallowF but with learnable number of units; Deep: deep neural network with up to 3 hidden layers.
Accel.: quantum reservoir accelerated (yes/no, individually investigated);
Activation function: relu, elu, tanh;
Units per layer: 16 to 256 (9 steps);
Dropout rate: 0.0 to 0.5 (6 steps);
Optimiser: Adam, Adagrad, RMSprop, SGD;
Learning rate: $10^{-5}$ to $10^{-1}$ (continuous);
Batch size: 32 to 256 (4 steps);
See Ref.~\cite{omalley2019kerastuner} for further details.
}
\begin{tabular}{lc|ccccccc}
\toprule
& & \multicolumn{7}{c}{Parameters} \\
\cmidrule(lr){3-9}
Model & Accel. & Hidden Layers & Activation Function & Units per Layer & Dropout Rate & Optimiser & Learning Rate & Batch Size \\
\midrule
Linear & \xmark & 0 & -- & -- & -- & Adam & $6.5 {\times} 10^{-4}$  & 256   \\
       & \cmark & 0 & -- & -- & -- & Adam & $2.2 {\times} 10^{-4}$  & 32   \\
ShallowF  & \xmark & 1 & relu & 16 & 0.0 & Adam    & $3.7 {\times} 10^{-4}$  & 32    \\
          & \cmark & 1 & relu & 16 & 0.0 & Adagrad & $2.0 {\times} 10^{-2}$ & 256 \\
Shallow & \xmark & 1 & elu  & 64 & 0.0 & RMSprop & $2.0 {\times} 10^{-3}$ & --     \\
        & \cmark & 1 & relu & 96 & 0.3 & RMSprop & $5.0 {\times} 10^{-4}$ & --     \\
Deep & \xmark & 3 & relu & $3{\times}$ 256 & 0.3/0.0/0.1 & Adam & $4.2 {\times} 10^{-4}$ & --  \\
     & \cmark & 3 & relu & 256/256/32  & 0.3/0.4/0.1 & Adam & $2.2 {\times} 10^{-4}$ & --  \\
\bottomrule
\end{tabular}
\end{table*}

\bibliographystyle{myapsrev4-2}
\bibliography{references}